\documentclass[useAMS]{mn2e}

\usepackage{graphicx}

\title[Fourier analysis of line profiles in rapidly rotating B-type stars]
{Fourier analysis of He~4471/Mg~4481 line profiles for separating 
rotational velocity and axial inclination in rapidly-rotating B-type stars\thanks{
Based on observations carried out at the Okayama Astronomical Observatory 
of National Astronomical Observatory of Japan.}
}

\author[Y. Takeda, S. Kawanomoto, and N. Ohishi]
{Y. Takeda$^{1}$\thanks{E-mail:
takeda.yoichi@nao.ac.jp}, 
S. Kawanomoto,$^{1}$ and N. Ohishi$^{1}$\\
$^{1}$National Astronomical Observatory of Japan, 
2-21-1 Osawa, Mitaka, Tokyo 181-8588, Japan\\
}
\begin{document}



\maketitle

\label{firstpage}

\begin{abstract}
While the effect of rotation on spectral lines is complicated 
in rapidly-rotating stars because of the appreciable gravity-darkening effect 
differing from line to line, it is possible to make use of this 
line-dependent complexity to separately determine the equatorial rotation
velocity ($v_{\rm e}$) and the inclination angle ($i$) of rotational axis.
Although line-widths of spectral lines were traditionally used for this aim,
we tried in this study to apply the Fourier method, which utilizes the 
unambiguously determinable first-zero frequency ($\sigma_{1}$) in 
the Fourier transform of line profile. 
Equipped with this technique, we analyzed the profiles of He~{\sc i} 4471 
and Mg~{\sc ii} 4481 lines of six rapidly-rotating 
($v_{\rm e}\sin i \sim$~150--300~km~s$^{-1}$) late B-type stars, 
while comparing them with the theoretical profiles simulated on 
a grid of models computed for various combination of ($v_{\rm e}$, $i$).
According to our calculation, $\sigma_{1}$ tends to be larger than the 
classical value for given $v_{\rm e}\sin i$. This excess progressively 
grows with an increase in $v_{\rm e}$, and is larger for the He line than 
the Mg line, which leads to $\sigma_{1}^{\rm He} > \sigma_{1}^{\rm Mg}$.
It was shown that $v_{\rm e}$ and $i$ are separately determinable from 
the intersection of two loci (sets of solutions reproducing the observed 
$\sigma_{1}$ for each line) on the $v_{\rm e}$ vs. $i$ plane. 
Yet, line profiles alone are not sufficient for their unique discrimination, 
for which photometric information (such as colors) needs to be 
simultaneously employed.
\end{abstract}


\section{Introduction}

In the widely used conventional treatment assuming the invariant line
profile over the disk along with the circular-symmetric brightness 
distribution of limb-darkening (valid for slow rotators), the effect of 
stellar rotation on a spectral line is quite simple, which can be universally 
expressed by convolution of the rotational broadening function with the 
intrinsic profile (see, e.g., Gray 2005). In this case, the equatorial rotational 
velocity ($v_{\rm e}$) and the projection factor ($\sin i$, where $i$ is the 
inclination angle of the rotation axis relative to the line of sight) always 
appears as both product ($v_{\rm e}\sin i$), and their separation is impossible.

Meanwhile, for the case of rapid rotators (e.g., $v_{\rm e} \sim$~100--400~km~s$^{-1}$) 
commonly seen in early-type stars, such an approximation is no more
valid in the practical sense, because the gravity-darkening effect 
(i.e., temperature inhomogeneity on the surface) causes an appreciable
variation in the strength and profile of spectral lines on the stellar disk, 
the extent of which is considerably line-dependent.          
Accordingly, the effect of rapid rotation on a spectral line is so complicated
as to be treated only by detailed numerical calculations based on a properly 
designed gravity-darkened stellar model. In compensation for this complexity,
however, separated determination of $v_{\rm e}$ and $i$ may be possible
by simultaneously analyzing the profiles of two or more lines of different
temperature sensitivity.

Actually, several investigators challenged this delicate and difficult task 
of $v_{\rm e}$--$i$ separation by carefully analyzing the spectral lines:
Stoeckley (1968a) studied five rapidly rotating B and A-type stars
by using He~{\sc i} 4471, Mg~{\sc ii} 4481, and Ca~{\sc ii} 3934 lines.
Hutchings \& Stoeckley (1977) examined the UV lines based on the Copernicus data 
(and also lines in the visual region) for 20 stars of mostly B-type.
Ruusalepp (1982) computed the half-width of He~{\sc i} 4471 and Mg~{\sc ii} 4481
lines for various model parameters and applied to 19 early B-type stars.
Stoeckley \& Buscombe (1987) extended the previous work of Stoeckley (1968a)
and analyzed the He~{\sc i} 4471 and Mg~{\sc ii} 4481 lines of 19 rapidly-rotating 
B-type main-sequence stars, where they also tried to detect the degree of 
differential rotation (in addition to independent determination of $v_{\rm e}$ and $i$).

Somewhat surprisingly, observational investigations of this kind seem to have 
been barely done since then over these 30 years, as far as we know,\footnote{
Vinicius, Townsend \& Leister (2007) reported their preliminary results
for the rotational inclination of five early B-type (Be) stars determined 
from the strengths (not the widths) of various spectral lines, though they 
do not appear to have published the details of their analysis.}  
When we review these old studies from the aspect of current knowledge,
some points are noticed, which may have room for further improvement.

First, rather rough interpolation appears to have been adopted in the 
simulation of spectral line profiles in gravity-darkened stellar photospheres, 
which requires integrations of local spectra at many points over the 
visible stellar disk. Presumably, this is due to the limitation of
computational capacity at the time. It may be possible nowadays to implement
more realistic computations; e.g., by using our code developed for simulating
the gravity-darkened spectrum of Vega (Takeda, Kawanomoto \& Ohishi 2008).

Second, these previous studies employed ``line widths'' (mostly half-width
at the half maximum of the line depth) for their analysis. However, precisely measuring 
the line width suffers practical difficulties for the relevant case of rapid rotators 
characterized by very wide and shallow profiles of merged line features, because it 
critically depends on the continuum level, for which exact determination is not easy. 
Given this situation, it occurred to us to apply the Fourier method, which utilizes 
the frequencies corresponding to the zero amplitude in the Fourier transform of a line 
profile, because  these zero-frequencies are easily determinable
from the amplitude vs. frequency diagram. 

It appears that this Fourier approach has not been so popularly applied 
to early-type stars, which is presumably related to the problem 
involved with rapid rotators commonly seen in hot stars. 
As well known, its merit for rotational velocity study is especially 
manifest when the emergent line flux profile ($F_{\lambda}$) is 
expressed by convolution of the intrinsic stellar flux profile 
($T_{\lambda}$) and the rotational broadening function ($G_{\lambda}$). 
That is, the value of $v_{\rm e}\sin i$ can be determined directly from 
the zeros of observed $f(\sigma)$ (transform of $F_{\lambda}$) without 
any necessity of deconvolving $T_{\lambda}$, because the zeros of 
$g(\sigma)$ (transform of $G_{\lambda}$) are automatically inherited to 
$f(\sigma)$ in the Fourier space where ``convolution'' in the wavelength 
space simply turns into ``multiplication.''
It should be remarked, however, that this conventional treatment
using the classical rotational broadening function 
[cf. Eq.~(18.14) in Gray (2005)] is based on the assumptions that 
(i) the line profile has the same shape at any point on the stellar 
disk and (ii) the brightness distribution on the stellar disk 
is circularly symmetric to be represented by a simple limb-darkening law,
which are evidently not valid any more for rapid rotators showing 
considerable gravity darkening as well as rotational distortion.
Actually, we know only a few trials of Fourier analysis for 
$v_{\rm e}\sin i$ determination with the classical rotational broadening 
function in the field of hot stars; e.g., Sim\'{o}n-D\'{\i}az \& Herrero 
(2007) for OB stars or D\'{\i}az et al. (2011) for A-type stars.
If line profiles of rapidly rotating stars are to be adequately analyzed, 
special considerations had to be made toward derivation of specific 
rotational broadening function, such as done by several investigators 
(cf. Stoeckley \& Mihalas 1973; Collins \& Truax 1995; Levenhagen 2014). 

Given this situation, we decided to carry out a feasibility study 
to examine whether it is possible to separately determine $v_{\rm e}$ 
and $i$ based on the Fourier analysis of line profiles for selected six 
rapidly-rotating late B-type stars, since such a trial seems to have 
been barely made to our knowledge.\footnote{
We note that several researchers have recently conducted 
sophisticated modeling of theoretical line profiles for 
rapidly-rotating B-type stars under the effect of gravity darkening, 
and also investigated the possibility of Fourier transform method 
for extracting in-depth information of stellar rotation 
such as the nature of differential rotation (e.g., Domiciano de Souza 
et al. 2004, Zorec et al. 2011, 2017). However, these studies placed 
emphasis on simulations of model profiles with particular intention 
of applying to Be stars based on combined interferometric and 
spectroscopic observations, which are thus somewhat different 
from the topic under question.}
For this purpose, we calculated a set of theoretical profiles on a grid 
of gravity-darkened models for various combinations of ($v_{\rm e}$, $i$). 
Here, we focus only on He~{\sc i} 4471 and Mg~{\sc ii} 4481 lines, 
which are suitable for the present aim and actually have been 
widely used so far.   
Our strategy is to compare the zeros in the Fourier transforms of the 
observed and the numerically simulated profiles directly to each other, 
In this sense, our analysis (except for the first preparatory analysis 
following the conventional procedure) is irrelevant to the concept of 
rotational broadening function (whichever classical or special) used to 
model the line profile by convolution. 
Yet, we also examine how the zeros in the Fourier transforms 
of the realistically computed line profiles of rapid rotators 
are compared with those of the classical rotational broadening, 
which would be worthwhile for evaluating the accuracies of apparent 
$v_{\rm e}\sin i$ values derived from the conventional treatment.

The remainder of this paper is organized as follows. Our observational data of six
target stars are described in Sect.~2. In Sect.~3 are presented the results of our 
preparatory analysis following the standard procedure (parameter determination
from colors, examination of positions on the HR diagram, spectrum-fitting analysis 
by using the classical rotational broadening function, 
evaluation of equivalent widths, etc.). Profile simulations of He~4471 and Mg~4481 
lines based on gravity-darkened models and Fourier transform calculations 
of observed as well as theoretical profiles are explained in Sect.~4. 
Sect.~5 is the discussion section, where the characteristics
of first-zero locations are examined, and we try separate determinations 
of $v_{\rm e}$ and $i$ by comparing the observed first-zero frequency with 
the theoretical grid. The summary of this investigation is given in Sect.~6.

\section{Observational Data}

As the targets of this study, we selected 6 apparently bright ($V \sim$~1--4)
late B-type stars (spectral type of B6--B9 and luminosity classes of III--V; 
cf. Table~1), all of which are known to be rapid rotators with 
$v_{\rm e}\sin i \sim$~100--300~km~s$^{-1}$ (e.g., Abt, Levato \& Grosso 2002). 

Note that 3 stars ($\eta$~Tau, $\beta$~CMi, 17~Tau) out of these 
6 program stars are classified as classical Be stars
according to Jaschek \& Egret's (1982) catalogue.
Actually, we can confirm the existence of appreciable emission component
in H$\alpha$ in the spectra of these 3 stars (especially for $\eta$~Tau 
and $\beta$~CMi).\footnote{
See, e.g., the spectral collection of Be stars available 
at http://www.astrosurf.com/buil/us/becat.htm .} 
Accordingly, we should keep in mind a possibility that the spectra may be  
contaminated by circumstellar Be disk for these stars. However, we 
consider that such an effect (even if any exists) is presumably not significant
for the He~4471/Mg~4481 lines, because no particular anomaly is seen 
in the strengths of these He and Mg lines for these Be stars as compared 
to those for other normal B stars (cf. Sect.~3.2).

The observations were carried out on 2006 November 1 ($\beta$~CMi, $\eta$~Tau), 
2 ($\zeta$~Peg), and 4 (17~Tau, $\alpha$~Leo, and $\eta$~Aqr) by using 
the HIgh-Dispersion Echelle Spectrograph (HIDES; Izumiura 1999) 
at the coud\'{e} focus of the 188~cm reflector of Okayama Astrophysical 
Observatory (OAO). Equipped with a 4K$\times$2K CCD detector at 
the camera focus, the HIDES spectrograph enabled us to obtain an 
echellogram covering a wavelength range of 3860--4630~$\rm\AA$ 
with a resolving power of $R \sim 70000$ (case for the normal slit 
width of 200~$\mu$m) in the mode of blue cross-disperser.
The total exposure time for each star ranged from 6~min to 100~min
depending on the stellar brightness or the weather condition.

The reduction of the spectra (bias subtraction, flat-fielding, 
scattered-light subtraction, spectrum extraction, wavelength 
calibration, co-addition of frames to increase the S/N ratio, 
and continuum normalization) was performed by using 
the ``echelle'' package of the software IRAF\footnote{
IRAF is distributed by the National Optical Astronomy Observatories,
which is operated by the Association of Universities for Research
in Astronomy, Inc. under cooperative agreement with the National 
Science Foundation.} in a standard manner. 
The S/N ratios of the finally resulting spectra turned out to be 
sufficiently high for the present purpose ($\ga$~500 for most stars 
at the 4450--4500~\AA\ region). 

\section{Conventional Model Atmosphere Analysis}

Before going into the Fourier analysis of spectral line profiles, 
we first conducted a preparatory study
following the conventional manner, in order to grasp the basic 
characteristics of program stars. 

\subsection{Stellar parameters}

The effective temperature ($T_{\rm eff}$) and the surface gravity 
($\log g$) of each star were determined from the colors of 
Str\"{o}mgren's $uvby\beta$ photometric system with the help of 
Napiwotzki, Sc\"{o}nberner, and Wenske's (1993) uvbybetanew
program\footnote{
$\langle$http://www.astro.le.ac.uk/\~{}rn38/uvbybeta.html$\rangle$.}, 
where the observational data of
$b-y$, $c_{1}$, $m_{1}$, and $\beta$ were taken from Hauck and 
Mermilliod (1998) via the SIMBAD database.
The resulting $T_{\rm eff}$ and $\log g$ are summarized in Table~1.

We also estimated the luminosity ($L$) of these stars by using the Hipparcos 
parallaxes (ESA 1997) along with the extinction correction (Arenou, 
Grenon \& Gomez 1992) and bolometric correction (Flower 1996), as done by 
Takeda et al. (2010). 
These $\log L$ values are plotted against $\log T_{\rm eff}$ 
in Fig.~1, where Ekstr\"{o}m et al.'s (2012) theoretical evolutionary tracks 
for both rotating and non-rotating models are also depicted. 
We can see from this figure that the masses of our sample 
stars are in the range between $\sim 3$~M$_{\odot}$ and $\sim 5$~M$_{\odot}$.
By comparing $M$ and $\log g$, we find that the radius values are $R\sim$~3--6~R$_{\odot}$
for most stars (excepting $\eta$~Tau, for which $R\sim$~10~R$_{\odot}$).

\subsection{Synthetic spectrum fitting}

Then, spectrum-synthesis analysis was carried out for the He~4471+Mg~4481 
line feature in the same manner as described in Sect.~4.2 of Takeda et al. (2010).
That is, while applying the optimization algorithm described in Takeda (1995),
we determined the solutions of $A$(He) (He abundance), $A$(Mg) (Mg abundance), 
$v_{\rm e}\sin i$ (projected rotational velocity), and $\Delta\lambda$
(radial velocity shift) accomplishing the best fit between the theoretical 
and observed spectra, where the rotational broadening was treated by convolving
the intrinsic flux profile with the rotational broadening function (limb-darkening 
coefficient of $\epsilon = 0.5$). Neither the instrumental broadening nor 
the macroturbulence broadening was taken into account, which are negligible
compared to the rotational broadening. 

The model atmosphere for each star used for this analysis was constructed
by two-dimensionally interpolating Kurucz's (1993) ATLAS9 model grid 
(solar-metallicity models) in terms of $T_{\rm eff}$ and $\log g$.
Regarding the He~{\sc i} 4471 line (multiplet No. 14; $\chi_{\rm low}$ = 20.96~eV), 
we adopted the total $gf$ value (summed over components) of 1.13 (cf. Table~2b 
of Takeda 1994) and the opacity profile (with forbidden components) 
broadened by Stark damping was treated according to Barnard, 
Cooper \& Smith (1974). As to the atomic line data for Mg~{\sc ii} 4481 
(Multiplet No. 4; $\chi_{\rm low}$ = 8.86~eV), we consulted the VALD database 
(Ryabchikova et al. 2015), which gives $\log gf$ = +0.740, $-0.560$, and +0.590 
for the components  at 4481.126, 4481.150, and 4481.325~\AA, respectively. 
The microturbulent velocity was fixed at 2~km~s$^{-1}$.

We assumed LTE throughout this study. The non-LTE effect for the He~4471 
and Mg~4481 lines is not likely to be important as far as 
late-B stars are concerned, even though it should act in the direction 
of intensifying the strengths of lines for both cases. 
According to Auer \& Mihalas (1973),
the non-LTE increase in the equivalent width for the He~4471 line
is only $\sim 5\%$ for the model of $T_{\rm eff} = 15000$~K and 
$\log g = 4.0$, which becomes even less significant as $T_{\rm eff}$
is lowered. Regarding the Mg~4481 line, the non-LTE abundance correction
for the A0V star Vega ($T_{\rm eff} \simeq 9500$~K and 
$\log g \simeq 4.0$)  was reported to be $\sim -0.1$~dex (Gigas 1988)
or $\sim -0.2$~dex (Przybilla et al. 2001), which corresponds to
a non-LTE intensification of equivalent width ($W \sim 300$~m\AA) 
by $\sim$~5--10\%. Since this non-LTE effect decreases with an 
increase of $T_{\rm eff}$ (i.e., with a decrease in $W$), it would
be insignificant in the regime of late B-type stars, which is
also supported by Mihalas's (1972) calculation for mid- through
early-B stars.

How the best-fit theoretical spectrum (corresponding to the converged parameter
solutions) matches the observed one for each star is shown in Fig.~2. 
It is instructive to note that an apparently satisfactory fit can be
achieved even by such a classical modeling using a plane-parallel
model atmosphere for given $T_{\rm eff}$ and $\log g$, if
$A$(He), $A$(Mg), $v_{\rm e}\sin i$, and $\Delta \lambda$ are adequately
adjusted. This means that separate determination of $v_{\rm e}$ and $i$ 
is a very delicate matter, which is hardly discernible by simply 
eye-inspecting the line profiles.
We also computed the equivalent widths $W$(He~4471) and $W$(Mg~4481) inversely 
from the converged solutions of $A$(He) and $A$(Mg) as done in Takeda et al. 
(2010; see Sect.~4.3 therein). The resulting values of $v_{\rm e}\sin i$,
$W$(He), and $W$(Mg) are presented in Table~1. 
Such derived $W$ values are plotted against $T_{\rm eff}$ in Fig.~3, where 
the theoretical relations are also depicted. We can see from this figure that
the trend of $W$ with a change in $T_{\rm eff}$ is roughly in agreement
with the theoretical prediction.

\section{Fourier Analysis of Spectral Lines}

\subsection{Transform of line profiles}

On both sides of a line for given flux spectrum $F(\lambda)$, we select two 
reference wavelengths ($\lambda_{1}$ and $\lambda_{2}$), at which we may 
regard $F(\lambda_{1})$ and $F(\lambda_{2})$ as the level of pseudo-continuum.  
The normalized line depth $R(\lambda)$ is defined within a limited
range of $\lambda_{1} \le \lambda \le \lambda_{2}$ as
\begin{equation}
R(\lambda) \equiv 1 - F(\lambda)/F_{\rm c} 
\end{equation} 
(while $R(\lambda)\equiv 0$ outside of this range), where $F_{\rm c}$ is 
the linear-interpolation of $F(\lambda_{1})$ and $F(\lambda_{2})$ in terms of $\lambda$:
\begin{equation}
F_{\rm c} \equiv \frac{\lambda_{2}-\lambda}{\lambda_{2}-\lambda_{1}}F(\lambda_{1})
               + \frac{\lambda-\lambda_{1}}{\lambda_{2}-\lambda_{1}}F(\lambda_{2}).
\end{equation}
The Fourier transform of $R(\lambda)$ is expressed as (see, e.g., Gray 2005)
\begin{equation}
  f(\sigma) \equiv \int_{-\infty}^{\infty} R(\lambda)\exp(2\pi i\sigma\lambda) {\rm d}\lambda.
\end{equation}

Regarding the choice of reference points in the observed spectrum, we paid attention 
to the critical wavelength ($\lambda^{*}$) of highest flux level between the He~4471 
and Mg~4481 lines (around $\sim$~4476--4478~\AA),
and equated as $\lambda_{2}^{\rm He}$ = $\lambda_{1}^{\rm Mg}$ = $\lambda^{*}$,
from which $\lambda_{1}^{\rm He}$ and $\lambda_{2}^{\rm Mg}$ were so chosen as 
to be symmetric relative to the line wavelength.
The relevant reference wavelengths ($\lambda_{1}^{\rm He}$, $\lambda^{*}$, and 
$\lambda_{2}^{\rm Mg}$) we adopted for each star are marked by green crosses in Fig.~2.

Fig.~4 illustrates the amplitudes of the computed Fourier transforms $|f(\sigma)|$ 
for the observed profiles of He~4471 and Mg~4481 lines as functions of $\sigma$
for each star.  We can see that the critical frequency ($\sigma_{1}$) corresponding 
to the first-zero of $f(\sigma)$ is clearly recognized by the cuspy feature and 
easily determinable. The measured values of $\sigma_{1}$ are given in Table~1.

\subsection{Spectrum simulation with gravity-darkened models}

Regarding the modeling of rotating stars with gravity darkening, we adopt
the same assumption and parameterization described in Sect. 2.1 and 2.2 of 
Takeda, Kawanomoto \& Ohishi (2008). Our model (Roche model with the assumption 
of rigid rotation) is characterized by stellar mass ($M$), polar radius ($R_{\rm p}$), 
polar effective temperature ($T_{\rm eff,p}$), and equatorial rotational velocity 
($v_{\rm e}$), which suffice to express the surface quantities [$r(\theta)$, 
$T_{\rm eff}(\theta)$, $g(\theta)$, $v(\theta)$] as functions of $\theta$ 
(colatitude angle).
Regarding the exponent $\beta$ in the relation $T_{\rm eff} \propto g^{\beta}$
determining the degree of gravity darkening, we used the $T_{\rm eff}$-dependent 
analytical formula [$\beta = f(T_{\rm eff})$]\footnote{
According to this formula, $\beta$ becomes equal to von Zeipel's value 
of 0.25 at the higher $T_{\rm eff}$ region of $\log T_{\rm eff} \ge 0.9$ 
($T_{\rm eff} \ge 7943$~K).
Since most of our models fall in this $T_{\rm eff}$ range (though 
the minimum $T_{\rm eff}$ is 7204~K in the exceptional case), 
we may say that the simple von Zeipel's law was adopted for 
almost all models.
However, we should keep in mind a possibility that actual $\beta$ may 
deviate from the von Zeipel value (0.25) even at such a higher $T_{\rm eff}$
range where the radiative equilibrium holds in the stellar envelope,
That is, according to Espinosa Lara \& Rieutord's (2011) calculation,
$\beta$ is a function of rotational flattening in the sense
that $\beta$ progressively decrease from $\sim 0.25$ (no rotation)
to $\sim 0.15$ (near to the rotational break-up limit). 
If this is really the case, the simple application of $\beta = 0.25$
to rapidly rotating stars may lead to an overestimation of gravity 
darkening.
} based on Fig.~5 of Claret (1998)
as done in Takeda, Kawanomoto \& Ohishi (2008; cf. footnote~2 therein).
Since different model atmosphere corresponding to the local ($T_{\rm eff}$, $g$) 
is defined at each point on the surface, we can compute the flux profile
to be observed by an external observer by integrating the emergent  
intensity profile over the visible disk for any inclination angle ($i$) of 
the rotational axis, while using the program CALSPEC (see Sect. 4.1 in 
Takeda, Kawanomoto \& Ohishi 2008).
As such, in order to compute a spectrum, it is necessary to specify five 
parameters: $M$, $R_{\rm p}$, $T_{\rm eff,p}$, $v_{\rm e}$, and $i$.
 
Given that our main purpose is to examine the feasibility of spectroscopically 
separating $v_{\rm e}$ and $i$, it is requisite to accomplish a sufficient
number of grid regarding these two parameters, while changing other parameters
is not necessarily important as long as their effects on line profiles are 
comparatively insignificant.
Therefore, considering the ranges of $M$ and $R$ estimated in Sect.~3.1, 
we assume $M$ = 4.0~M$_{\odot}$ and $R$ = 4.0~R$_{\odot}$ as fixed.
Since the photometric $T_{\rm eff}$ derived from colors ($\sim$~11000-13000~K; 
cf. Table~1) should reflect the mean $T_{\rm eff}$ averaged over the visible disk, 
the polar $T_{\rm eff,p}$ must be appreciably larger, though the difference
intricately depends upon $v_{\rm e}$ and $i$. We tentatively assume two 
$T_{\rm eff,p}$  cases of 12000~K and 15000~K, hoping that they sufficiently 
cover the temperature range of 6 program stars.
Accordingly, adopting the same line data and the assumptions as described 
in Sect.~3.2 along with the solar abundances of He and Mg, we computed 120 
($ = 6 \times 10 \times 2$) spectra of He~4471+Mg~4481 line feature, 
corresponding to combinations of 6 $v_{\rm e}$ values (100, 150, 200, 250, 300, 
and 350~km~$^{-1}$), 10 $i$ values (0, 10, 20, 30, 40, 50, 60, 70, 80, and 
90$^{\circ}$), and 2 $T_{\rm eff,p}$ values (12000~K and 15000~K).
The parameters of these models are listed in Table~2, where
the corresponding values of $\langle T_{\rm eff} \rangle$ (intensity-weighted
average of $T_{\rm eff}$ over the visible disk) are also presented.   
Several selected examples of the resulting spectra are depicted in Fig.~5. 

The wavelengths of the reference pseudo-continuum, necessary for reducing the
normalized profile, were defined in almost the same way as in Sect.~4.1:
Having found the critical wavelength $\lambda^{*}$ corresponding to the maximum level 
between these two lines, we set  [$\lambda_{1}^{\rm He}$, $\lambda_{2}^{\rm He}$] = 
[$4471 - (\lambda^{*} - 4471)$, $\lambda^{*}$] and
[$\lambda_{1}^{\rm Mg}$, $\lambda_{2}^{\rm Mg}$] = 
[$\lambda^{*}$,  $4481+ (4481 - \lambda^{*})$] .
The Fourier transforms were then calculated for the resulting $R(\lambda)$ profiles 
of He~4471 and Mg~4481 lines. Fig.~6 illustrates the behaviors of $|f(\sigma)|$ 
as function of $\sigma$ for the same selected cases as in Fig.~5. 
The first-zero frequencies ($\sigma_{1}$) measured from the computed transforms 
are presented in Table 2. 

\section{Discussion}

\subsection{Characteristics of first-zero frequencies}

We first discuss the trend of theoretical $\sigma_{1}$ modeled in Sect.~4.2.
Since this quantity is expressed as 
$\sigma_{1} = 0.660/(\lambda v_{\rm e}\sin i/c)$ in the classical approximation,
the projected rotational velocity (or the line-width for the case of 
rotation-dominated broadening) plays the most significant role,
though its precise value is affected also by the profile shape.
The resulting values of $\sigma_{1}$ corresponding to the theoretical line profiles 
simulated on the gravity-darkened models are plotted against $v_{\rm e}\sin i$
in Fig.~7, where the classical $\sigma_{1}$ vs. $v_{\rm e}\sin i$ relation is 
depicted by a slanted solid line and the observed $\sigma_{1}$ values 
for 6 stars are indicated by horizontal dashed lines.  

We can see from this figure that the $\sigma_{1}$ values of the simulated 
spectra closely follow the classical relation (i.e., inversely proportional 
to $v_{\rm e}\sin i$), as long as $v_{\rm e}\sin i$ is small
(e.g., up to several tens km~s$^{-1}$), $\sigma_{1}$, though appreciable 
deviations ($\sigma_{1}$ tends to be larger than the classical value) 
reflecting the gravity-darkening effect begin to be recognized
as $v_{\rm e}\sin i$ becomes larger. A closer inspection of the figure
reveals that (i) the extent of deviation is essentially determined by $v_{\rm e}$,
(ii) this discrepancy is more manifest for He~4471 than for Mg~4481, and
(iii) no significant difference exists regarding the trend of $\sigma_{1}$
between the $T_{\rm eff,p}$ = 12000~K and 15000~K cases (though the deviation
being slightly larger for the former).

The fact that $\sigma_{1}$ tends to be in excess of the classical value
is reasonably explained by considering the physical effect caused by
gravity darkening. In the rotationally-broadened spectral lines, the line 
width is mainly determined by maximum Doppler velocity caused by
rotation (i.e., $v_{\rm e}\sin i$), to which the low-latitude 
region near to the equator makes most significant contribution
because the absolute rotational velocity is largest there.
In the case of rapid rotators with large $v_{\rm e}$, these equatorial 
regions are considerably darkened and thus their contribution to line 
broadening becomes lessened. Accordingly, the width becomes 
narrower compared to the classical case, which eventually shifts 
$\sigma_{1}$ to a larger value. Moreover, this effect is further amplified
for the case of He~4471, because the strength of this line quickly
drops as $T_{\rm eff}$ is lowered (cf. Fig.~3). Consequently,
the inequality relation $\sigma_{1}^{\rm He} > \sigma_{1}^{\rm Mg}$
holds for rapid-rotator models, which is actually observed
in our program stars (cf. Fig.~4).

Another important implication read from Fig.~7 is that $\sigma_{1}$
tends to be stabilized at the high-rotation limit ($\ga 300$~km~s$^{-1}$),
in the sense that $\sigma_{1}$ does not effectively decrease 
(or line width does not effectively increase) any more
no matter how the rotational velocity is increased, which is because
two effects (increase of rotation and enhanced gravity darkening) 
on the width counteract with each other and tend to be cancelled. 
Actually, this effect was first pointed out by Stoeckley (1968b) and also 
confirmed by Townsend, Owocki \& Howarth (2004); i.e., rotational velocities 
of rapidly-rotating stars (especially those close to the limit) may be 
significantly underestimated if simply determined from the apparent line widths.  

\subsection{Separation of rotation and inclination}

Now that we have theoretical $\sigma_{1}$ values as functions 
of $v_{\rm e}$ and $i$ for He~4471 as well as Mg~4481 lines, we can 
compare them with the observed $\sigma_{1}$.  
Then, a possible set of ($v_{\rm e}$, $i$) solutions is expressed  
as a locus on the $v_{\rm e}$ vs. $i$ plane by solving the equation
$\sigma_{1}^{\rm th}(v_{\rm e}, i) = \sigma_{1}^{\rm obs}$ for each line,
from which ($v_{\rm e}, i$) may be established from the intersection of 
two loci on the plane. Such constructed loci (displayed in Fig.~8), 
and the estimation of ($v_{\rm e}, i$) values (summarized in Table~1) 
are discussed below star by star.
 
\subsubsection{$\zeta$~Peg}

Fig.~8a and Fig.~8a' represent the characteristic difficulty involved with 
this analysis. Since the loci derived for He and Mg lines are rather similar 
in shape, the intersection is not clearly defined. To our embarrassment,
this similarity brings about a marked difference of ($v_{\rm e}, i$)
solution depending on $T_{\rm eff,p}$, despite that $\sigma_{1}$ is
not so sensitive to it (cf. Sect.~4.2). That is, 
($\sim$150--200~km~s$^{-1}$, $\sim$50--60$^{\circ}$) for 12000~K and 
($\sim$250--300~km~s$^{-1}$, $\sim$30--40$^{\circ}$) for 15000~K,
Fortunately, however, the former (more equator-on-like slower rotation)  
and the latter (more pole-on-like faster rotation) yield distinctly
different $\langle T_{\rm eff} \rangle$; i.e., $\sim$11200--11500~K
and $\sim$13000--13700~K according to Table~2. Comparing these with 
$T_{\rm eff}^{\rm color}$ of $\sim$12700~K, we can guess that the 
actual solution would be almost the intermediate between these 
two as ($\sim$200--250~km~s$^{-1}$, $\sim$40--50$^{\circ}$).
Stoeckley \& Buscombe (1987) derived [195 (158--246)~km~s$^{-1}$, 
54 (40--90)$^{\circ}$] for this star, nearly consistent with our result. 

\subsubsection{$\eta$~Aqr}

This star has the largest $v_{\rm e}\sin i$ of $\sim 290$~km~s$^{-1}$
among our 6 sample stars. Unfortunately, we were unable find the solution
for this star, because the $\sigma_{1}^{\rm obs}$ values (0.172 for He and 
0.153 for Mg) are too small to be covered by the grid of $\sigma_{1}^{\rm th}$ 
(see Fig.~7). The reason for this incompatibility is not clear.
It might be possible that the physical condition can not be adequately 
described by our modeling for such case of very rapid rotator.
Alternatively, there might be systematic errors in the measurement of $\sigma_{1}$
for such case of especially large $v_{\rm e} \sin i$, because He and Mg lines 
tend to somewhat merge in-between (Fig.~2).
In this context, the ambiguity in the measurement of $\sigma_{1}^{\rm He}$ would be 
larger than $\sigma_{1}^{\rm Mg}$, because the cuspy feature near to the zero 
amplitude is less sharp for the former than the latter (cf. Fig.~4b).  
As a test, we increased the original $\sigma_{1}^{\rm He}$ arbitrarily by 10\% 
($0.172 \rightarrow 0.189$) and made a retry. In this case, we could find 
a solution for the $T_{\rm eff,p}$ = 15000~K case (but failed again for
the 12000~K case) as shown in Fig.~8b', which suggests ($\sim$300~km~s$^{-1}$, 
$\sim$80--90$^{\circ}$). In any event, we may conclude that this star is a very 
rapidly rotating star ($v_{\rm e}\sim$300~km~s$^{-1}$ or even somewhat larger) seen 
nearly equator-on. In such cases, solution search becomes especially difficult, 
because $i = 90^{\circ}$ is a singularity point, in the sense that it can not be
encompassed by $i$ values of the grid. As such, the feasibility of establishing 
the solution is quite vulnerable to errors in $\sigma_{1}$ or modeling inadequacies.
  
\subsubsection{$\eta$~Tau}

The ($v_{\rm e}, i$) for this star appears to be comparatively easier to embrace.
Since the intersection of He and Mg loci lies at ($\sim$300~km~s$^{-1}$, $\sim$35$^{\circ}$)
for $T_{\rm eff,p}$ = 12000~K (Fig.~8c) and  at ($\sim$350~km~s$^{-1}$, $\sim$30$^{\circ}$)
for $T_{\rm eff,p}$ = 15000~K (Fig.~8c') and not much different from each other,
we may regard that this star is a very rapid rotator with 
$v_{\rm e}\sim$300--350~km~s$^{-1}$ seen from a rather low inclination angle 
($\i \sim$30--35$^{\circ}$). 

\subsubsection{$\beta$~CMi}

We could not find a reasonable solution for this star. First, $\sigma_{1}^{\rm He}$
is so small (as is the case for $\eta$~Aqr) and no He-locus exists for 
$T_{\rm eff,p}$ = 12000~K (Fig.~8d). Meanwhile,  no clear intersection point 
can be defined for the $T_{\rm eff,p}$ = 15000~K case (Fig.~8d').
Although we may roughly speculate from these figures $v_{\rm e} \sim$~250--300~km~s$^{-1}$ 
and $i \sim$~60--80$^{\circ}$ as the possible ranges (i.e., rather high inclination angle), 
any conclusion should be withholded. On the other hand, Stoeckley (1968a) reported
(as a tentative conclusion) the aspect angle of this star to be $i \sim$30--50$^{\circ}$. 
 
\subsubsection{$\alpha$~Leo}

Almost the same situation holds for this star as the case of $\eta$~Aqr,
since the $\sigma_{1}$ values are indiscernibly similar to each other (cf. Fig.~7),
which means that no solution exists because $\sigma_{1}$ is outside of the range 
of the grid. Again, we increased $\sigma_{1}^{\rm He}$ by 10\% ($0.172 \rightarrow 0.189$) , 
as a trial and searched for the ($v_{\rm e}$, $i$) solution. Naturally, the results 
are essentially the same as the case for $\eta$~Aqr (compare Fig.~8e' with Fig.~8b'),
suggesting ($\sim$300~km~s$^{-1}$, $\sim$80--90$^{\circ}$) as a rough estimate.
We may thus at least state that $\alpha$~Leo is a very rapidly-rotating star 
seen nearly equator-on. This is consistent with the previous determinations:
(300~km~s$^{-1}$, 90$^{\circ}$) estimated by Hutchings \& Stoeckley (1977) 
from the widths of UV lines, [285 (270--309)~km~s$^{-1}$, 90 (67--90)$^{\circ}$] 
determined by Stoeckley \& Buscombe (1987) from the widths of He~4471 and Mg~4481 lines,
and [$317 \pm 3$~km~s$^{-1}$, 75--90$^{\circ}$] concluded by McAlister et al. (2005) 
based on interferometric observations.

\subsubsection{17~Tau}

The behavior of loci on the $v_{\rm e}$ vs. $i$ plane for this star is similar
to the case of $\zeta$~Peg. The intersection of He and Mg loci depends
on $T_{\rm eff,p}$ as ($\sim$200~km~s$^{-1}$, $\sim$60$^{\circ}$)
for 12000~K and ($\sim$300~km~s$^{-1}$, $\sim$30--40$^{\circ}$)
for 15000~K, each of which correspond to $\langle T_{\rm eff} \rangle$
of $\sim$11000~K and $\sim$13000~K according to Table~2.
Considering that $T_{\rm eff}^{color}$ is $\sim$12700~K, we conclude that 
the latter solution is relevant, which means that this star is very 
rapidly-rotating with $v_{\rm e} \sim 300$~km~s$^{-1}$ and seen with a 
rather low aspect angle ($i \sim$~30--40$^{\circ}$). 

\subsubsection{Comparison with Zorec et al.'s (2016) results}

Zorec et al. (2016) recently studied the rotational velocity distribution of
a large sample of 233 Be stars. They also separated rotation and inclination 
for each star by comparing the observed stellar parameters affected by rotation 
(effective temperature, surface gravity, luminosity, projected rotational velocity) 
with extensive theoretical calculations based on gravity-darkened stellar models 
(cf. Appendix E therein). 
Since three of our targets are included in their sample,
it is interesting to compare our profile-based results 
of axial inclination with their determinations.
Comparing their $i$ values ($62\pm 15^{\circ}$, $66\pm 16^{\circ}$, 
and $45\pm 11^{\circ}$ for $\eta$~Tau, $\beta$~CMi, and 17~Tau, 
respectively) with our results presented in Table~1 
($\sim$30--35$^{\circ}$, $\sim$60--80$^{\circ}$, and
$\sim$30--40$^{\circ}$), we notice an appreciable disagreement 
for $\eta$~Tau, although a reasonable consistency is seen for 
$\beta$~CMi and 17~Tau.
This discrepancy for $\eta$~Tau may be attributed to the fact 
that this is an exceptional case of evolved subgiant ($R\sim 10$R$_{\odot}$) 
among our sample (cf. Sect.~3.2). Our result for this star 
would be less reliable, because theoretical calculations 
corresponding to main-sequence B stars ($R=4$R$_{\odot}$) were 
applied to it like others. 

\section{Summary and Conclusion}

It is known that the gravity-darkening effect causes an appreciable 
latitudinal inhomogeneity on the surface of a rapidly-rotating star; 
i.e., lower $T_{\rm eff}$ as well as $g$ at lower latitude (in short, 
cool/dark equator and hot/bright pole).
Owing to this effect, the impact of rotation on the shape of spectral lines
is complicated and different from line to line, and the simple 
classical treatment (convolution of the intrinsic profile with the rotational 
broadening function determined by $v_{\rm e}\sin i$) is no more valid.  

From the converse point of view, it is possible to make advantage of this 
line-dependent complexity to separately determine the equatorial rotational 
velocity ($v_{\rm e}$) and the inclination angle ($i$) of rotational axis, 
which is impossible within the framework of the conventional approximation.
Although several investigators challenged this task several decades ago,
those old studies appear rather outdated as viewed from the present-day 
standard, especially in terms of their policy of simply invoking line-widths 
and seemingly insufficient accuracy in simulating line-profiles.

We thus tried to examine in this study whether the Fourier method,
which is a comparatively modern technique utilizing the unambiguously 
determinable first-zero frequency of the Fourier transform of a line profile, 
is applicable to this problem of spectroscopically separating $v_{\rm e}$ and $i$. 
Conveniently, since we already have a computer code of simulating line profiles 
for a rapidly-rotating star with distorted surface of inhomogeneous brightness, 
we can make use of it.
As to the lines to be analyzed, we chose He~{\sc i} 4471 and Mg~{\sc ii} 4481 
lines, which have been often used for this purpose because they are strong 
and have markedly different temperature sensitivity. 

Toward this aim, six rapidly-rotating late B-type stars
($\zeta$~Peg, $\eta$~Aqr, $\eta$~Tau, $\beta$~CMi, $\alpha$~Leo, 17~Tau) 
were selected as our targets, for which high-dispersion spectra of
sufficient quality are available. We first evaluated $T_{\rm eff}$ and $\log g$ 
from $uvby\beta$ colors, and then carried out a conventional spectrum-fitting 
analysis on the He~4471+Mg~4481 line feature based on plane-parallel model 
atmospheres to estimate $v_{\rm e}\sin i$ and the equivalent widths of two lines.

The theoretical line profiles of gravity-darkened rapid rotators were simulated for 
a grid of models for various combinations of $v_{\rm e}$ (100--350~km~s$^{-1}$)
and $i$ (0--90$^{\circ}$), while typical values were assumed for other 
parameters ($T_{\rm eff,p}$ = 12000~K and 15000~K, $M_{\rm p}$ = 4.0~M$_{\odot}$, 
$R_{\rm p}$ = 4.0~R$_{\odot}$). We then computed their Fourier transforms and
measured the frequencies corresponding to the first zero ($\sigma_{1}$).

These modeled $\sigma_{1}$ values revealed the characteristic trends,
reflecting the gravity-darkening effect and the different
temperature susceptibility of these lines. 
(i) They tend to be in excess of the classical value
$\sigma_{1}^{\rm cl} = 0.660 /(\lambda v_{\rm e}\sin i/c)$
when compared at a given $v_{\rm e}\sin i$, and this difference progressively
grows with an increase in $v_{\rm e}$.  
(ii) The amount of this excess is larger for He~4471 than for Mg~4481, resulting 
in an inequality relation $\sigma_{1}^{\rm He} > \sigma_{1}^{\rm Mg}$.

Then, by comparing the $\sigma_{1}$ of the observed profile with the theoretical 
grid of $\sigma_{1}(v_{\rm e}, i)$ for each of the He~4471 and Mg~4481 line, 
we can define two loci on the $v_{\rm e}$ vs. $i$ plane, from which ($v_{\rm e}, i$)
may be separately determined from their intersection.
We tried this solution search for 6 program stars, and confirmed that the intersection 
point is measurable, except for the difficult case of largest $v_{\rm e}\sin i$ 
($\eta$~Aqr and $\alpha$~Leo; equator-on case with considerably large $v_{\rm e}$).
However, since the position of this intersection tends to depend upon $T_{\rm eff,p}$, 
its adequate specification was often necessarily, for which we compared 
$\langle T_{\rm eff} \rangle$ of the theoretical grid
with $T_{\rm eff}^{\rm color}$. That is, line profiles alone are not   
sufficient but photometric information needs to be simultaneously employed.

To conclude, we could show in this investigation that the Fourier method 
using the first-zero frequencies in the transforms of line profiles 
is effectively applicable to separation of $v_{\rm e}$ and $i$. 
Although this approach is not essentially different from the traditional 
method using line widths, it has a merit that characteristic zero point of 
the transform is comparatively easier to measure even for the case of rapid rotators.

Yet, such a rough feasibility study as attempted in this paper still 
has room for further improvements. We enumerate below several issues 
to be considered in the future. 
\begin{itemize}

\item
Although we employed only He~4471 and Mg~4481 lines,
simultaneously using more lines of different parameter sensitivity 
would surely improve the reliability of the solutions.

\item
Our adopted grid of theoretical gravity-darkened models was rather rough, 
especially in terms of the parameters other than $v_{\rm e}$ and $i$
(i.e., $T_{\rm eff,p}$, $R_{\rm p}$), for which much finer mesh would be required. 

\item
Since the assumption of rigid rotation is not necessarily guaranteed,
additional parameter describing the degree of differential rotation 
would desirably be included.

\item
Similarly, the possibility of rotation-dependence in the exponent 
$\beta$ (see footnote~6), which determines the degree of gravity 
darkening, should be seriously investigated.

\item
In our Fourier analysis of line profiles, we made use of only 
$\sigma_{1}$, which is nothing but one of the various quantities 
characterizing the Fourier transform. It may be possible to utilize 
other observables such as the second-zero frequency or the amplitude of 
the side lobe. 
\end{itemize}
When advanced analysis has been carried out by adequately taking 
account of these points, more realistic and trustworthy results 
would be expected. 

\section*{Acknowledgments}

This research has made use of the SIMBAD database, operated by CDS, Strasbourg, France. 
This work has also made use of the VALD database, operated at Uppsala 
University, the Institute of Astronomy RAS in Moscow, and the University of Vienna.

\newpage

\setcounter{table}{0}
\begin{table*}
\begin{minipage}{130mm}
\scriptsize
\caption{Basic parameters and observed data of program stars.}
\begin{center}
\begin{tabular}{ccccccccccccc}\hline
Name & HR\# & HD\# & Sp.Type & $V$ & $T_{\rm eff}^{\rm color}$ & $\log g^{\rm color}$ & $v_{\rm e}\sin i$ &
$W_{4471}^{\rm He}$ & $W_{4481}^{\rm Mg}$ & $\sigma_{1}^{\rm He}$ & $\sigma_{1}^{\rm Mg}$ & ($v_{\rm e}$, $i$) \\
(1) & (2) & (3) & (4) & (5) & (6) & (7) & (8) & (9) & (10) & (11) & (12) & (13)\\
\hline
$\zeta$~Peg   &8634 & 214923 &  B8~V    &3.41 & 11182 & 3.65 & 153 & 346 & 329 & 0.305 & 0.283 & ($\sim$200--250, $\sim$40--50)\\
$\eta$~Aqr    &8597 & 213998 &  B9~IV-Vn &4.03 & 11458 & 3.91 & 289 & 450 & 357 & 0.172$^{*}$ & 0.153 & ($\sim$300, $\sim$80--90)\\
$\eta$~Tau    &1165 &  23630 &  B7~III   &2.87 & 11599 & 2.50 & 158 & 546 & 280 & 0.340 & 0.269 & ($\sim$300--350, $\sim$30--35)\\
$\beta$~CMi   &2845 &  58715 &  B8~Ve    &2.89 & 11696 & 3.42 & 231 & 431 & 323 & 0.203 & 0.191 & ($\sim$250--300, $\sim$60--80)\\
$\alpha$~Leo  &3982 &  87901 &  B8~IVn   &1.40 & 12223 & 3.54 & 276 & 511 & 274 & 0.172$^{*}$ & 0.152 & ($\sim$300, $\sim$80--90)\\
17~Tau        &1142 &  23302 &  B6~IIIe  &3.70 & 12698 & 3.28 & 164 & 851 & 275 & 0.287 & 0.254 & ($\sim$300, $\sim$30--40)\\
\hline
\end{tabular}
\end{center}
(1) Star name. (2) HR number. (3) HD number. (4) Spectral type from SIMBAD. (5) Apparent visual magnitude from SIMBAD (in mag).
(6) Effective temperature from $uvby\beta$ (in K). (7) Logarithmic surface gravity from $uvby\beta$ (in cm~s$^{-1}$).
(from $uvby\beta$). (8) Fitting-based projected rotational velocity (in km~s$^{-1}$). (9) Fitting-based equivalent 
width of He~{\sc i} 4471 (in m\AA). (10) Fitting-based equivalent width of Mg~{\sc i} 4481 (in m\AA). 
(11) First-zero frequency of He~{\sc i} 4471 (in $\rm\AA^{-1}$).  
(12) First-zero frequency of Mg~{\sc ii} 4481(in $\rm\AA^{-1}$). 
(13) Estimated solution of ($v_{\rm e}$, $i$), where $v_{\rm e}$ is in km~s$^{-1}$
and $i$ is in degree (see Sect.~5.2 for the details).\\
$^{*}$Actually, an arbitrarily increased value by 10\% (0.189) was used for the 
($v_{\rm e}$, $i$) solution search.
\end{minipage}
\end{table*}

\setcounter{table}{1}
\begin{table*}
\begin{minipage}{180mm}
\scriptsize
\caption{Computed models of rotating stars and the first-zero frequencies of Fourier transforms.}
\begin{center}
\begin{tabular}{ccccccccccc}\hline
Code & $R_{\rm p}$ & $R_{\rm e}$ & $T_{\rm eff,p}$ & $T_{\rm eff,e}$ & $\langle T_{\rm eff} \rangle$ & 
$v_{\rm e}\sin i$ & $\sigma_{1}^{\rm He}$ & $\sigma_{1}^{\rm He}/\sigma_{1}^{\rm cl}$  & $\sigma_{1}^{\rm Mg}$ & 
$\sigma_{1}^{\rm Mg}/\sigma_{1}^{\rm cl}$ \\
(1) & (2) & (3) & (4) & (5) & (6) & (7) & (8) & (9) & (10) & (11) \\
\hline
m40r40t120v100i00 &4.00 &4.11 &12000 &11679 &11854 &   0.0 &   1.938& $\cdots$ &   2.304& $\cdots$ \\
m40r40t120v100i10 &     &     &      &      &11852 &  17.4 &   2.481&   0.974 &   2.297&   0.903 \\
m40r40t120v100i20 &     &     &      &      &11844 &  34.2 &   1.308&   1.011 &   1.287&   0.997 \\
m40r40t120v100i30 &     &     &      &      &11831 &  50.0 &   0.895&   1.011 &   0.877&   0.994 \\
m40r40t120v100i40 &     &     &      &      &11815 &  64.3 &   0.712&   1.034 &   0.682&   0.993 \\
m40r40t120v100i50 &     &     &      &      &11798 &  76.6 &   0.603&   1.043 &   0.573&   0.993 \\
m40r40t120v100i60 &     &     &      &      &11781 &  86.6 &   0.535&   1.047 &   0.505&   0.991 \\
m40r40t120v100i70 &     &     &      &      &11768 &  94.0 &   0.503&   1.068 &   0.471&   1.002 \\
m40r40t120v100i80 &     &     &      &      &11759 &  98.5 &   0.479&   1.065 &   0.444&   0.991 \\
m40r40t120v100i90 &     &     &      &      &11755 & 100.0 &   0.472&   1.068 &   0.439&   0.993 \\
\hline
m40r40t120v150i00 &4.00 &4.25 &12000 &11257 &11673 &   0.0 &   3.012& $\cdots$ &   2.304& $\cdots$ \\
m40r40t120v150i10 &     &     &      &      &11667 &  26.0 &   1.688&   0.993 &   1.702&   1.004 \\
m40r40t120v150i20 &     &     &      &      &11650 &  51.3 &   0.900&   1.043 &   0.868&   1.008 \\
m40r40t120v150i30 &     &     &      &      &11622 &  75.0 &   0.634&   1.075 &   0.592&   1.006 \\
m40r40t120v150i40 &     &     &      &      &11586 &  96.4 &   0.503&   1.096 &   0.456&   0.997 \\
m40r40t120v150i50 &     &     &      &      &11546 & 114.9 &   0.424&   1.100 &   0.384&   0.999 \\
m40r40t120v150i60 &     &     &      &      &11507 & 129.9 &   0.364&   1.068 &   0.342&   1.007 \\
m40r40t120v150i70 &     &     &      &      &11473 & 141.0 &   0.344&   1.096 &   0.310&   0.989 \\
m40r40t120v150i80 &     &     &      &      &11450 & 147.7 &   0.321&   1.072 &   0.299&   0.998 \\
m40r40t120v150i90 &     &     &      &      &11442 & 150.0 &   0.315&   1.068 &   0.292&   0.991 \\
\hline
m40r40t120v200i00 &4.00 &4.47 &12000 &10620 &11421 &   0.0 &   1.938& $\cdots$ &   2.304& $\cdots$ \\
m40r40t120v200i10 &     &     &      &      &11411 &  34.7 &   1.358&   1.065 &   1.295&   1.018 \\
m40r40t120v200i20 &     &     &      &      &11383 &  68.4 &   0.723&   1.117 &   0.661&   1.023 \\
m40r40t120v200i30 &     &     &      &      &11336 & 100.0 &   0.498&   1.125 &   0.453&   1.026 \\
m40r40t120v200i40 &     &     &      &      &11274 & 128.6 &   0.393&   1.143 &   0.347&   1.011 \\
m40r40t120v200i50 &     &     &      &      &11200 & 153.2 &   0.321&   1.112 &   0.293&   1.018 \\
m40r40t120v200i60 &     &     &      &      &11123 & 173.2 &   0.285&   1.115 &   0.255&   1.000 \\
m40r40t120v200i70 &     &     &      &      &11055 & 187.9 &   0.270&   1.148 &   0.237&   1.011 \\
m40r40t120v200i80 &     &     &      &      &11007 & 197.0 &   0.258&   1.147 &   0.225&   1.002 \\
m40r40t120v200i90 &     &     &      &      &10990 & 200.0 &   0.249&   1.127 &   0.227&   1.027 \\
\hline
m40r40t120v250i00 &4.00 &4.78 &12000 & 9689 &11101 &   0.0 &   3.018& $\cdots$ &   2.299& $\cdots$ \\
m40r40t120v250i10 &     &     &      &      &11089 &  43.4 &   1.156&   1.134 &   1.057&   1.039 \\
m40r40t120v250i20 &     &     &      &      &11051 &  85.5 &   0.616&   1.189 &   0.538&   1.043 \\
m40r40t120v250i30 &     &     &      &      &10986 & 125.0 &   0.425&   1.199 &   0.367&   1.038 \\
m40r40t120v250i40 &     &     &      &      &10893 & 160.7 &   0.338&   1.228 &   0.288&   1.049 \\
m40r40t120v250i50 &     &     &      &      &10775 & 191.5 &   0.279&   1.206 &   0.237&   1.029 \\
m40r40t120v250i60 &     &     &      &      &10644 & 216.5 &   0.249&   1.220 &   0.215&   1.053 \\
m40r40t120v250i70 &     &     &      &      &10520 & 234.9 &   0.231&   1.229 &   0.195&   1.038 \\
m40r40t120v250i80 &     &     &      &      &10429 & 246.2 &   0.227&   1.265 &   0.191&   1.063 \\
m40r40t120v250i90 &     &     &      &      &10395 & 250.0 &   0.219&   1.238 &   0.182&   1.032 \\
\hline
m40r40t120v300i00 &4.00 &5.24 &12000 & 8247 &10730 &   0.0 &   1.932& $\cdots$ &   2.310& $\cdots$ \\
m40r40t120v300i10 &     &     &      &      &10718 &  52.1 &   1.045&   1.230 &   0.910&   1.074 \\
m40r40t120v300i20 &     &     &      &      &10682 & 102.6 &   0.566&   1.312 &   0.456&   1.060 \\
m40r40t120v300i30 &     &     &      &      &10614 & 150.0 &   0.400&   1.354 &   0.319&   1.083 \\
m40r40t120v300i40 &     &     &      &      &10506 & 192.8 &   0.314&   1.367 &   0.243&   1.061 \\
m40r40t120v300i50 &     &     &      &      &10351 & 229.8 &   0.261&   1.355 &   0.207&   1.079 \\
m40r40t120v300i60 &     &     &      &      &10155 & 259.8 &   0.238&   1.396 &   0.182&   1.070 \\
m40r40t120v300i70 &     &     &      &      & 9949 & 281.9 &   0.225&   1.433 &   0.170&   1.085 \\
m40r40t120v300i80 &     &     &      &      & 9787 & 295.4 &   0.227&   1.515 &   0.164&   1.098 \\
m40r40t120v300i90 &     &     &      &      & 9725 & 300.0 &   0.225&   1.524 &   0.166&   1.125 \\
\hline
m40r40t120v350i00 &4.00 &5.89 &12000 & 7204 &10375 &   0.0 &   1.926& $\cdots$ &   2.316& $\cdots$ \\
m40r40t120v350i10 &     &     &      &      &10370 &  60.8 &   1.016&   1.395 &   0.830&   1.142 \\
m40r40t120v350i20 &     &     &      &      &10352 & 119.7 &   0.548&   1.481 &   0.421&   1.141 \\
m40r40t120v350i30 &     &     &      &      &10316 & 175.0 &   0.387&   1.532 &   0.286&   1.132 \\
m40r40t120v350i40 &     &     &      &      &10250 & 225.0 &   0.308&   1.564 &   0.225&   1.147 \\
m40r40t120v350i50 &     &     &      &      &10126 & 268.1 &   0.255&   1.544 &   0.197&   1.193 \\
m40r40t120v350i60 &     &     &      &      & 9936 & 303.1 &   0.239&   1.640 &   0.178&   1.222 \\
m40r40t120v350i70 &     &     &      &      & 9705 & 328.9 &   0.234&   1.737 &   0.164&   1.219 \\
m40r40t120v350i80 &     &     &      &      & 9504 & 344.7 &   0.230&   1.795 &   0.166&   1.296 \\
m40r40t120v350i90 &     &     &      &      & 9422 & 350.0 &   0.236&   1.870 &   0.166&   1.319 \\
\hline
\end{tabular}
\end{center}
\end{minipage}
\end{table*}

\setcounter{table}{1}
\begin{table*}
\begin{minipage}{180mm}
\scriptsize
\caption{(Continued.)}
\begin{center}
\begin{tabular}{ccccccccccc}\hline
Code & $R_{\rm p}$ & $R_{\rm e}$ & $T_{\rm eff,p}$ & $T_{\rm eff,e}$ & $\langle T_{\rm eff} \rangle$ & 
$v_{\rm e}\sin i$ & $\sigma_{1}^{\rm He}$ & $\sigma_{1}^{\rm He}/\sigma_{1}^{\rm cl}$  & $\sigma_{1}^{\rm Mg}$ & 
$\sigma_{1}^{\rm Mg}/\sigma_{1}^{\rm cl}$ \\
(1) & (2) & (3) & (4) & (5) & (6) & (7) & (8) & (9) & (10) & (11) \\
\hline
m40r40t150v100i00 &4.00 &4.11 &15000 &14599 &14816 &   0.0 &   1.932& $\cdots$ &   2.329& $\cdots$ \\
m40r40t150v100i10 &     &     &      &      &14812 &  17.4 &   2.413&   0.947 &   2.347&   0.923 \\
m40r40t150v100i20 &     &     &      &      &14802 &  34.2 &   1.241&   0.959 &   1.281&   0.992 \\
m40r40t150v100i30 &     &     &      &      &14787 &  50.0 &   0.852&   0.962 &   0.880&   0.996 \\
m40r40t150v100i40 &     &     &      &      &14768 &  64.3 &   0.675&   0.981 &   0.684&   0.996 \\
m40r40t150v100i50 &     &     &      &      &14747 &  76.6 &   0.575&   0.996 &   0.575&   0.997 \\
m40r40t150v100i60 &     &     &      &      &14727 &  86.6 &   0.504&   0.987 &   0.507&   0.995 \\
m40r40t150v100i70 &     &     &      &      &14710 &  94.0 &   0.472&   1.003 &   0.463&   0.984 \\
m40r40t150v100i80 &     &     &      &      &14699 &  98.5 &   0.443&   0.986 &   0.446&   0.995 \\
m40r40t150v100i90 &     &     &      &      &14695 & 100.0 &   0.437&   0.987 &   0.440&   0.997 \\
\hline
m40r40t150v150i00 &4.00 &4.25 &15000 &14071 &14585 &   0.0 &   1.944& $\cdots$ &   2.329& $\cdots$ \\
m40r40t150v150i10 &     &     &      &      &14578 &  26.0 &   1.602&   0.943 &   1.690&   0.997 \\
m40r40t150v150i20 &     &     &      &      &14556 &  51.3 &   0.839&   0.973 &   0.853&   0.991 \\
m40r40t150v150i30 &     &     &      &      &14523 &  75.0 &   0.593&   1.005 &   0.585&   0.994 \\
m40r40t150v150i40 &     &     &      &      &14479 &  96.4 &   0.467&   1.016 &   0.459&   1.003 \\
m40r40t150v150i50 &     &     &      &      &14431 & 114.9 &   0.393&   1.021 &   0.386&   1.004 \\
m40r40t150v150i60 &     &     &      &      &14383 & 129.9 &   0.344&   1.011 &   0.334&   0.983 \\
m40r40t150v150i70 &     &     &      &      &14342 & 141.0 &   0.320&   1.018 &   0.310&   0.991 \\
m40r40t150v150i80 &     &     &      &      &14314 & 147.7 &   0.297&   0.991 &   0.300&   1.003 \\
m40r40t150v150i90 &     &     &      &      &14304 & 150.0 &   0.291&   0.986 &   0.292&   0.993 \\
\hline
m40r40t150v200i00 &4.00 &4.47 &15000 &13275 &14262 &   0.0 &   1.944& $\cdots$ &   2.334& $\cdots$ \\
m40r40t150v200i10 &     &     &      &      &14250 &  34.7 &   1.243&   0.975 &   1.277&   1.004 \\
m40r40t150v200i20 &     &     &      &      &14215 &  68.4 &   0.657&   1.016 &   0.648&   1.004 \\
m40r40t150v200i30 &     &     &      &      &14158 & 100.0 &   0.461&   1.040 &   0.438&   0.993 \\
m40r40t150v200i40 &     &     &      &      &14082 & 128.6 &   0.357&   1.036 &   0.341&   0.992 \\
m40r40t150v200i50 &     &     &      &      &13992 & 153.2 &   0.301&   1.044 &   0.286&   0.992 \\
m40r40t150v200i60 &     &     &      &      &13899 & 173.2 &   0.260&   1.018 &   0.257&   1.008 \\
m40r40t150v200i70 &     &     &      &      &13817 & 187.9 &   0.246&   1.043 &   0.239&   1.019 \\
m40r40t150v200i80 &     &     &      &      &13760 & 197.0 &   0.231&   1.030 &   0.226&   1.010 \\
m40r40t150v200i90 &     &     &      &      &13740 & 200.0 &   0.233&   1.055 &   0.219&   0.990 \\
\hline
m40r40t150v250i00 &4.00 &4.78 &15000 &12111 &13847 &   0.0 &   1.944& $\cdots$ &   2.323& $\cdots$ \\
m40r40t150v250i10 &     &     &      &      &13832 &  43.4 &   1.012&   0.992 &   1.025&   1.008 \\
m40r40t150v250i20 &     &     &      &      &13785 &  85.5 &   0.557&   1.077 &   0.518&   1.003 \\
m40r40t150v250i30 &     &     &      &      &13705 & 125.0 &   0.376&   1.061 &   0.353&   0.999 \\
m40r40t150v250i40 &     &     &      &      &13590 & 160.7 &   0.290&   1.055 &   0.273&   0.995 \\
m40r40t150v250i50 &     &     &      &      &13447 & 191.5 &   0.252&   1.091 &   0.231&   1.002 \\
m40r40t150v250i60 &     &     &      &      &13288 & 216.5 &   0.218&   1.067 &   0.208&   1.020 \\
m40r40t150v250i70 &     &     &      &      &13139 & 234.9 &   0.209&   1.110 &   0.188&   1.000 \\
m40r40t150v250i80 &     &     &      &      &13031 & 246.2 &   0.194&   1.079 &   0.183&   1.019 \\
m40r40t150v250i90 &     &     &      &      &12992 & 250.0 &   0.194&   1.097 &   0.185&   1.045 \\
\hline
m40r40t150v300i00 &4.00 &5.24 &15000 &10309 &13347 &   0.0 &   1.938& $\cdots$ &   2.328& $\cdots$ \\
m40r40t150v300i10 &     &     &      &      &13332 &  52.1 &   0.906&   1.067 &   0.865&   1.021 \\
m40r40t150v300i20 &     &     &      &      &13284 & 102.6 &   0.492&   1.140 &   0.439&   1.020 \\
m40r40t150v300i30 &     &     &      &      &13196 & 150.0 &   0.333&   1.129 &   0.298&   1.011 \\
m40r40t150v300i40 &     &     &      &      &13057 & 192.8 &   0.260&   1.134 &   0.239&   1.045 \\
m40r40t150v300i50 &     &     &      &      &12860 & 229.8 &   0.225&   1.167 &   0.203&   1.054 \\
m40r40t150v300i60 &     &     &      &      &12618 & 259.8 &   0.200&   1.174 &   0.176&   1.036 \\
m40r40t150v300i70 &     &     &      &      &12370 & 281.9 &   0.188&   1.197 &   0.164&   1.046 \\
m40r40t150v300i80 &     &     &      &      &12180 & 295.4 &   0.191&   1.273 &   0.158&   1.057 \\
m40r40t150v300i90 &     &     &      &      &12108 & 300.0 &   0.182&   1.232 &   0.160&   1.084 \\
\hline
m40r40t150v350i00 &4.00 &5.89 &15000 & 7384 &12835 &   0.0 &   1.938& $\cdots$ &   2.328& $\cdots$ \\
m40r40t150v350i10 &     &     &      &      &12828 &  60.8 &   0.876&   1.203 &   0.780&   1.074 \\
m40r40t150v350i20 &     &     &      &      &12807 & 119.7 &   0.467&   1.264 &   0.396&   1.075 \\
m40r40t150v350i30 &     &     &      &      &12766 & 175.0 &   0.326&   1.290 &   0.276&   1.093 \\
m40r40t150v350i40 &     &     &      &      &12683 & 225.0 &   0.258&   1.314 &   0.213&   1.084 \\
m40r40t150v350i50 &     &     &      &      &12517 & 268.1 &   0.221&   1.342 &   0.182&   1.106 \\
m40r40t150v350i60 &     &     &      &      &12264 & 303.1 &   0.203&   1.389 &   0.172&   1.183 \\
m40r40t150v350i70 &     &     &      &      &11964 & 328.9 &   0.197&   1.465 &   0.158&   1.175 \\
m40r40t150v350i80 &     &     &      &      &11710 & 344.7 &   0.193&   1.505 &   0.160&   1.249 \\
m40r40t150v350i90 &     &     &      &      &11608 & 350.0 &   0.193&   1.530 &   0.151&   1.199 \\
\hline
\end{tabular}
\end{center}
(1) Model code (m**r**t***v***i**), which describes the adopted stellar mass ($M$), polar radius
($R_{\rm p}$), polar effective temperature ($T_{\rm eff,p}$), equatorial rotation velocity ($v_{\rm e}$),
and inclination angle ($i$). For example, ``m40r40t150v300i30'' is the case of
$M = 4.0{\rm M}_{\odot}$, $R_{\rm p} = 4.0{\rm R}_{\odot}$, $T_{\rm eff,p} = 15000$~K, 
$v_{\rm e} = 300$~km~s$^{-1}$, and $i = 30^{\circ}$. (2) Polar radius (in R$_{\odot}$).
(3) Equatorial radius (in R$_{\odot}$). (4) Polar effective temperature (in K).
(5) Equatorial effective temperature (in K). (6) Mean effective temperature in K (which was derived 
by averaging the local effective temperature over the observed stellar disk while weighting 
according to the local continuum intensity). (7) Product of $v_{\rm e}$ and $\sin i$ 
(in km~s$^{-1}$). (8) First-zero frequency of He~4471 (in $\rm\AA^{-1}$). 
(9) Ratio of $\sigma_{1}^{\rm He}$ to $\sigma_{1}^{\rm cl}$,
where $\sigma_{1}^{\rm cl}$ is the first-zero frequency of classical
rotational broadening function (see, e.g., Gray 2005), which is related to $v_{\rm e}\sin i$ as
$\sigma_{1}^{\rm cl} = 0.660/(\lambda v_{\rm e}\sin i / c) $ ($\lambda$: wavelength,
$c$: velocity of light). (10) First-zero frequency of Mg~4481 (in $\rm\AA^{-1}$).
(11) Ratio of $\sigma_{1}^{\rm Mg}$ to $\sigma_{1}^{\rm cl}$.
\end{minipage}
\end{table*}

\newpage

\setcounter{figure}{0}
\begin{figure*}
\begin{minipage}{90mm}
\includegraphics[width=9.0cm]{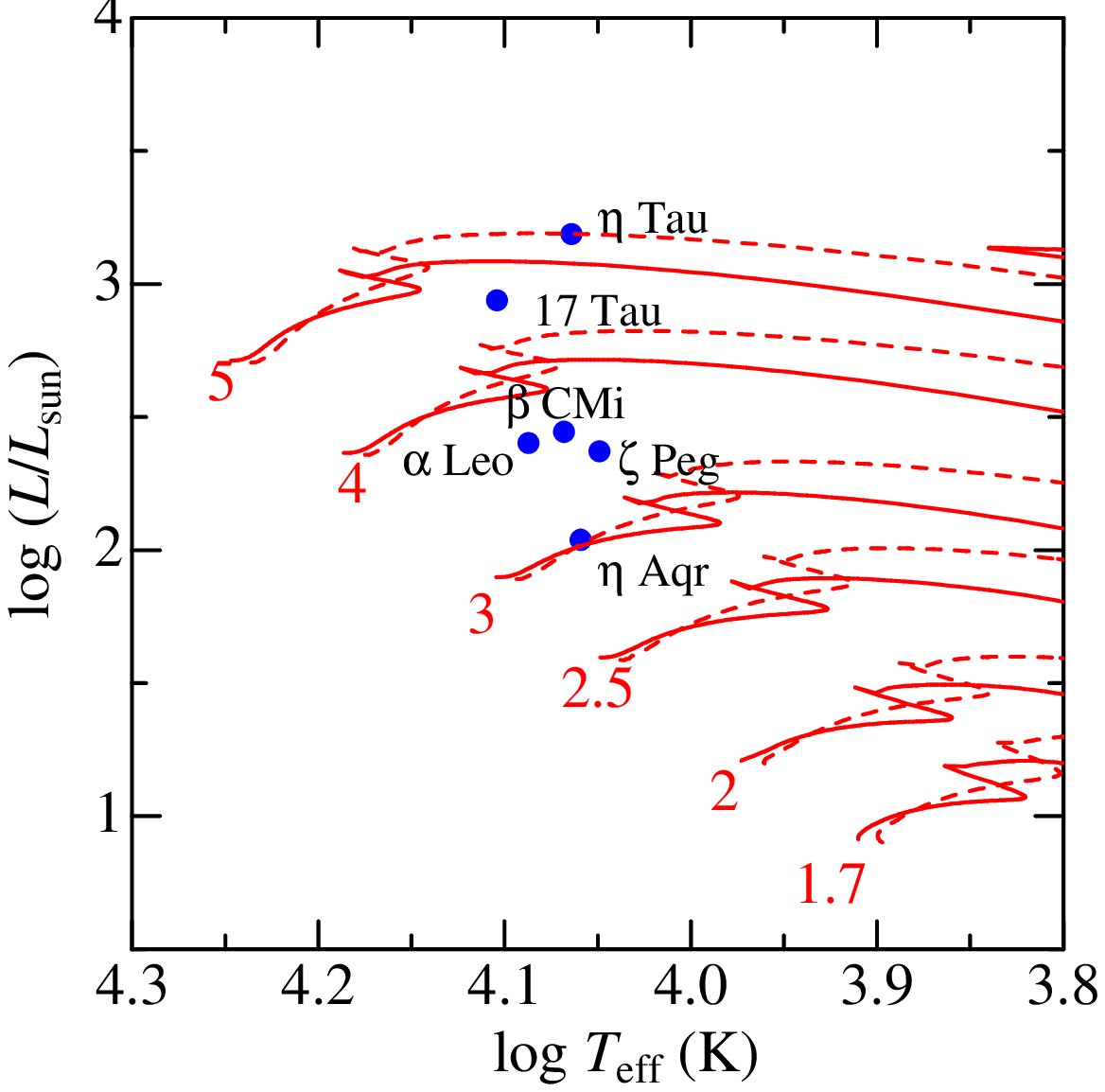}
\caption{
Plots of the 6 program stars on the $\log (L/{\rm L}_{\odot})$ vs. $\log T_{\rm eff}$
diagram, where the effective temperature ($T_{\rm eff}$) was determined from 
$uvby\beta$ photometry (Sect.~2) and the bolometric luminosity ($L$) was 
evaluated from the apparent visual magnitude (Table~1), Hipparcos parallax 
(ESA 1997), Arenou, Grenon \& G\'{o}mez's (1992) interstellar extinction
map, and Flower's (1996) bolometric correction. 
Theoretical evolutionary tracks corresponding to the solar metallicity 
computed by Ekstr\"{o}m et al. (2012) for six different initial masses 
(1.7, 2, 2.5, 3, 4, and 5~M$_{\odot}$) are also depicted for comparison,
where solid lines and dashed lines correspond to the non-rotating models and 
the rotating models (with 40\% of the critical break-up velocity), respectively.
}
\label{fig1}
\end{minipage}
\end{figure*}

\setcounter{figure}{1}
\begin{figure*}
\begin{minipage}{130mm}
\includegraphics[width=13.0cm]{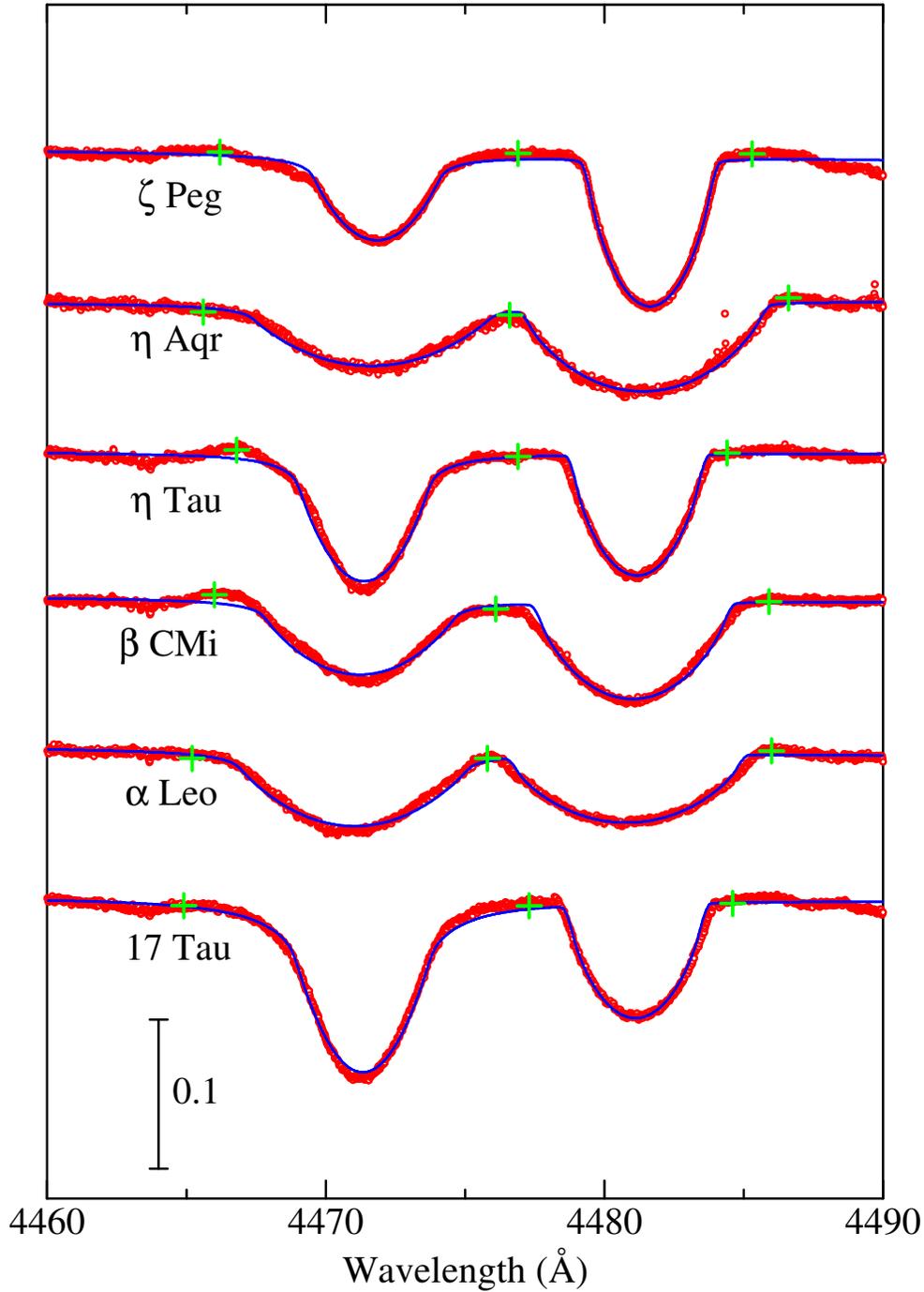}
\caption{
Synthetic spectrum fitting in the 4460--4490~\AA\ region comprising
He~{\sc i} 4471 and Mg~{\sc ii} 4481 lines,
which was accomplished based on the conventional modeling (using the plane-parallel
model atmosphere and the classical rotational broadening function)
by adjusting four parameters [$A$(He), $A$(Mg), $v_{\rm e}\sin i$, 
$\Delta \lambda$; cf. Sect.~3.2]. 
The observed data are plotted by red symbols, while the best-fit 
theoretical spectra are shown by blue solid lines. 
The spectra are arranged (from top to bottom) in the ascending order 
of $T_{\rm eff}$ as in Table~1, and a vertical offset of 0.1 is applied 
to each spectrum relative to the adjacent one. 
The green crosses indicate the reference points of pseudo-continuum
(cf. Sect. 4.1), based on which the normalized line-depth profiles 
(to be used for calculating the Fourier transforms) were computed. 
}
\label{fig2}
\end{minipage}
\end{figure*}

\setcounter{figure}{2}
\begin{figure*}
\begin{minipage}{100mm}
\includegraphics[width=10.0cm]{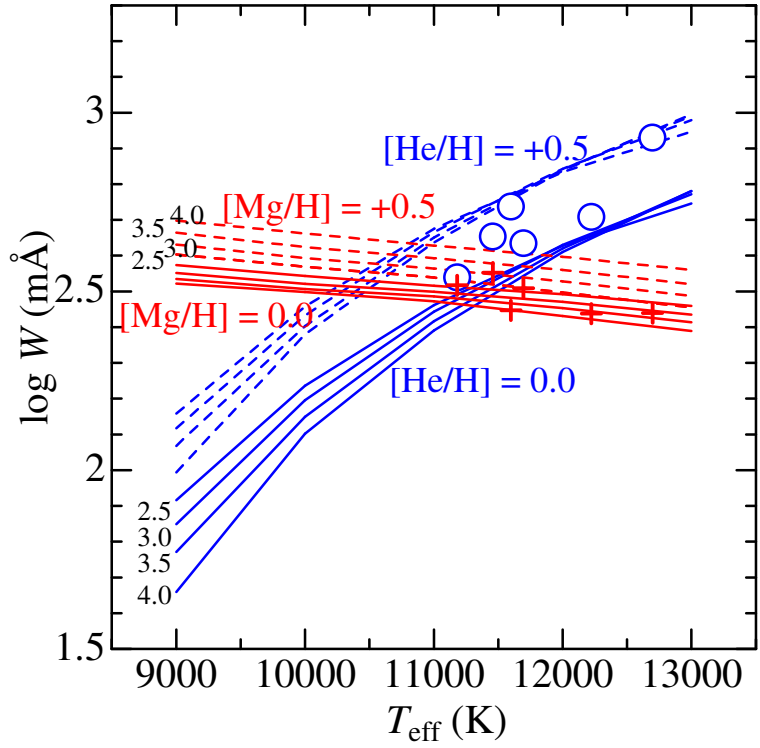}
\caption{
Equivalent widths ($W$) of He~{\sc i} 4471 (blue) and Mg~{\sc ii} (red) 
plotted against the effective temperature. The theoretical $W$ values (computed for 
models with $T_{\rm eff}$ = 9000, 10000, 11000, 12000, and 13000~K, and $\log g$ = 
2.5, 3.0, 3.5, and 4.0) are depicted by lines (solid lines: solar abundance, 
dashed lines: overabundance by +0.5~dex), while the observed values for 6 program stars 
are shown by symbols (open circles for He 4471 and crosses for Mg 4481). 
Note the markedly large dependence of $W$(He) upon $T_{\rm eff}$ (increasing with 
$T_{\rm eff}$), while $W$(Mg) is only weakly $T_{\rm eff}$-dependent (decreasing
with $T_{\rm eff}$). Regarding the gravity dependence, $W$(He) tends to decrease 
with an increase in $\log g$ (indicated in the figure), while the tendency is 
reversed for $W$(Mg).   
}
\label{fig3}
\end{minipage}
\end{figure*}

\setcounter{figure}{3}
\begin{figure*}
\begin{minipage}{150mm}
\includegraphics[width=15.0cm]{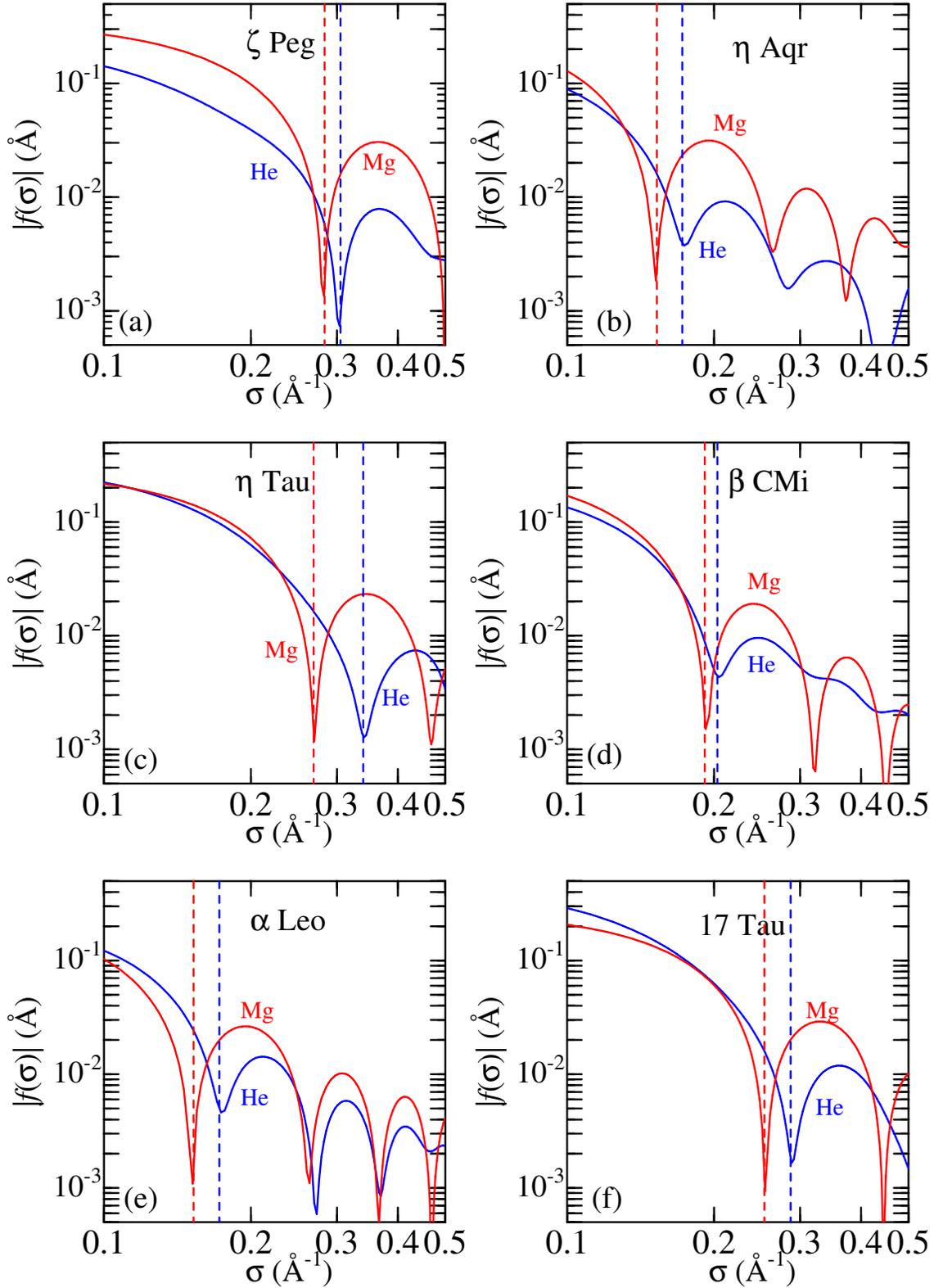}
\caption{
Fourier transform amplitudes [$|f(\sigma)|$] plotted against $\sigma$, 
for the observed profiles of He~4471 (blue) and Mg~4481 (red).
The positions of the first zero ($\sigma_{1}$) are indicated by vertical dashed lines.
(a) $\zeta$~Peg, (b) $\eta$~Aqr, (c) $\eta$~Tau, (d) $\beta$~CMi,
(e) $\alpha$~Leo, and (e) 17~Tau. 
}
\label{fig4}
\end{minipage}
\end{figure*}

\setcounter{figure}{4}
\begin{figure*}
\begin{minipage}{120mm}
\includegraphics[width=12.0cm]{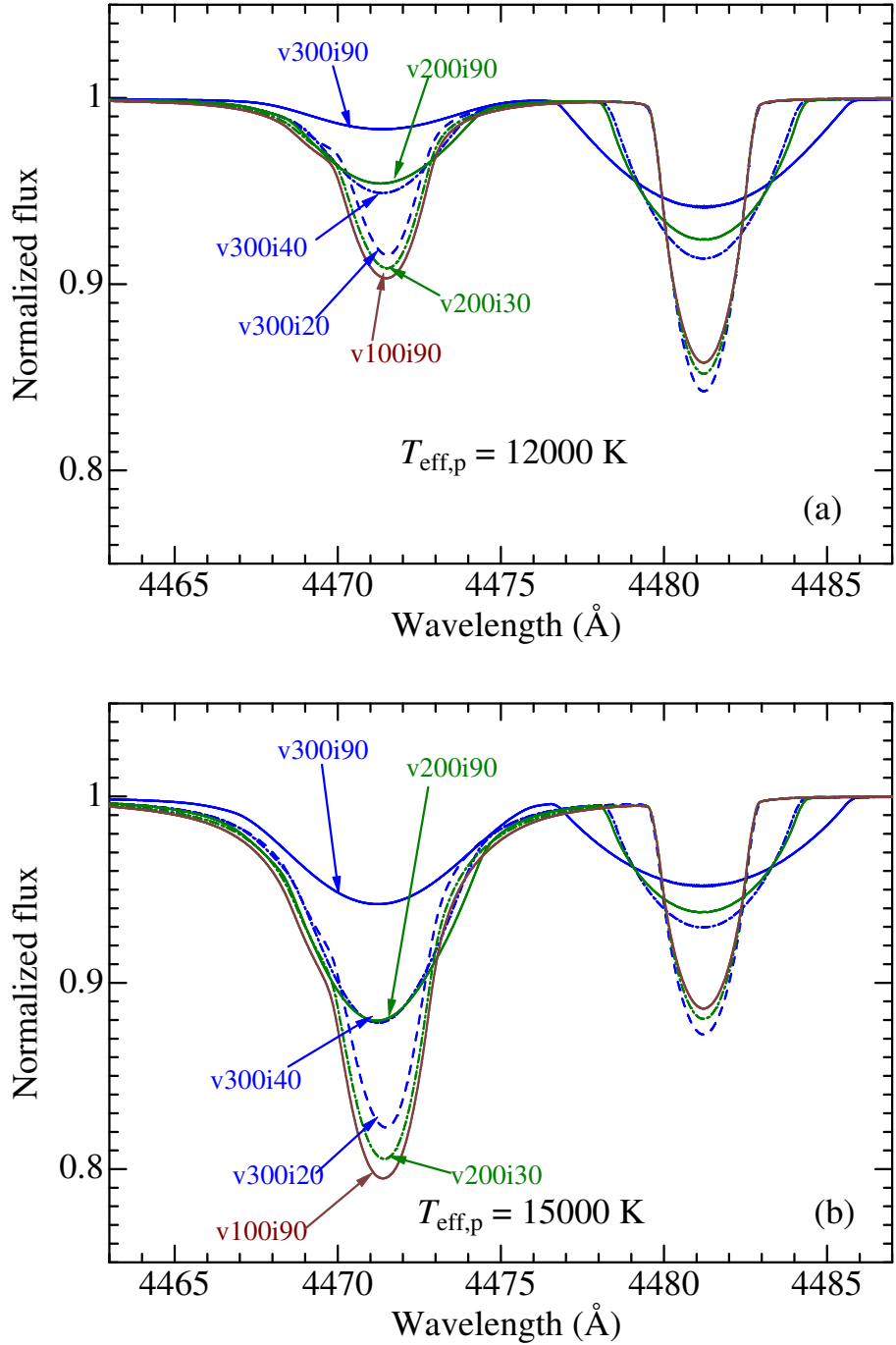}
\caption{
Theoretically synthesized  spectra in the 4463--4487~\AA\ region computed for 
six representative models of v100i90, v200i30, v200i90, v300i20, v300i40, 
and v300i90 (the code ``v***i**'' means $v_{\rm e}$ = *** km~s$^{-1}$ and
$i$ = **$^{\circ}$), which were so chosen as to demonstrate the
three $v_{\rm e}\sin i$ cases of $\sim 100$, $\sim 200$, and
300~km~s$^{-1}$. The results for $v_{\rm e}$ = 100, 200, and 300~km~s$^{-1}$
are depicted in brown, green, and blue, respectively.
The upper panel (a) shows the case for $T_{\rm eff,p}$ = 12000~K,
while the lower panel (b) is for $T_{\rm eff,p}$ = 15000~K.
}
\label{fig5}
\end{minipage}
\end{figure*}

\setcounter{figure}{5}
\begin{figure*}
\begin{minipage}{150mm}
\begin{center}
\includegraphics[width=15.0cm]{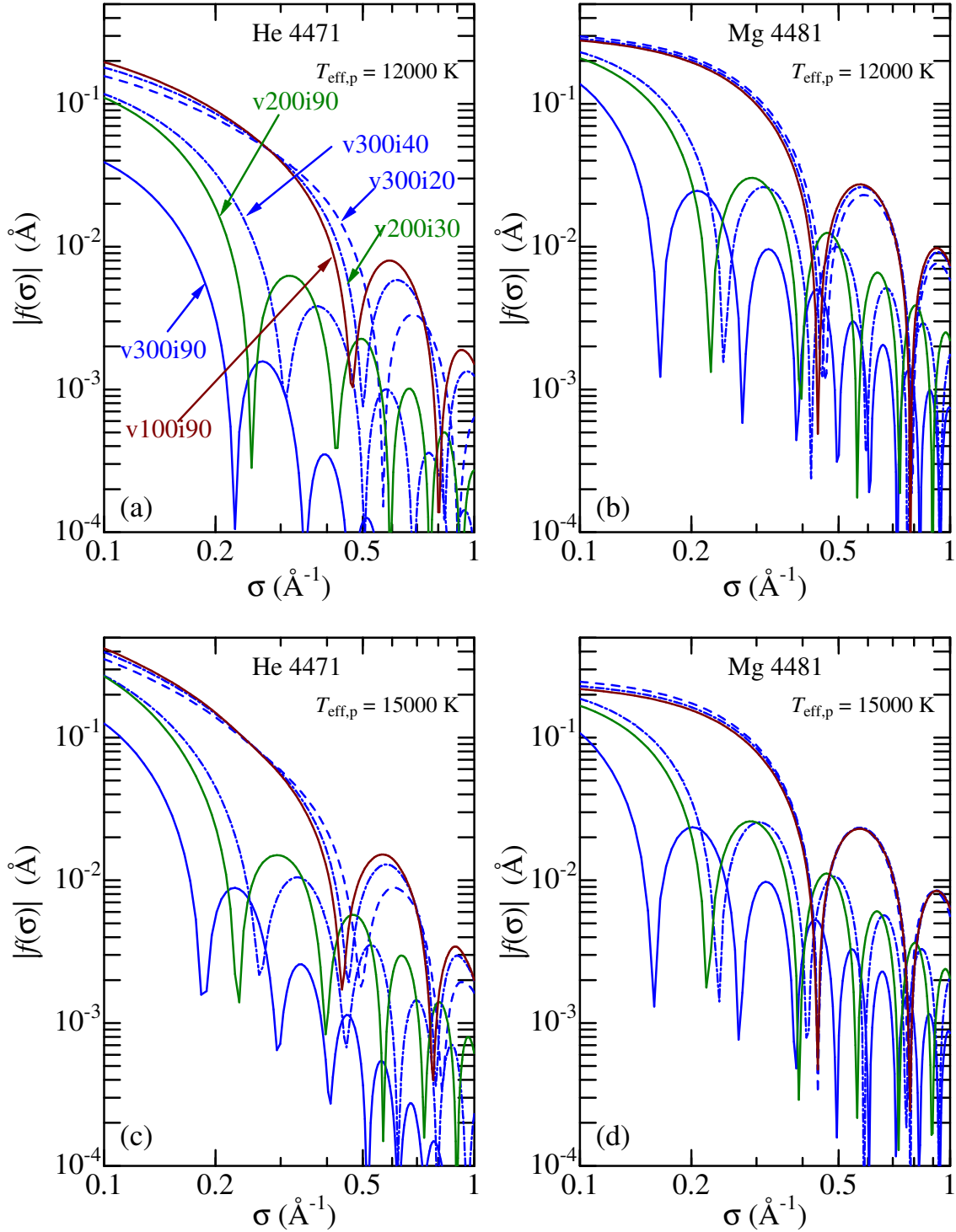}
\caption{
Fourier transform amplitudes [$|f(\sigma)|$] plotted against $\sigma$,
for the theoretical profiles of He~4471 (left panels) and Mg~4481 (right panels) 
computed for six representative models (same as those shown in Fig.~5).
The upper panels (a) and (b) show the results for $T_{\rm eff,p}$ = 12000~K,
while the lower panels (c) and (d) are for $T_{\rm eff,p}$ = 15000~K.
}
\label{fig6}
\end{center}
\end{minipage}
\end{figure*}

\setcounter{figure}{6}
\begin{figure*}
\begin{minipage}{140mm}
\begin{center}
\includegraphics[width=14.0cm]{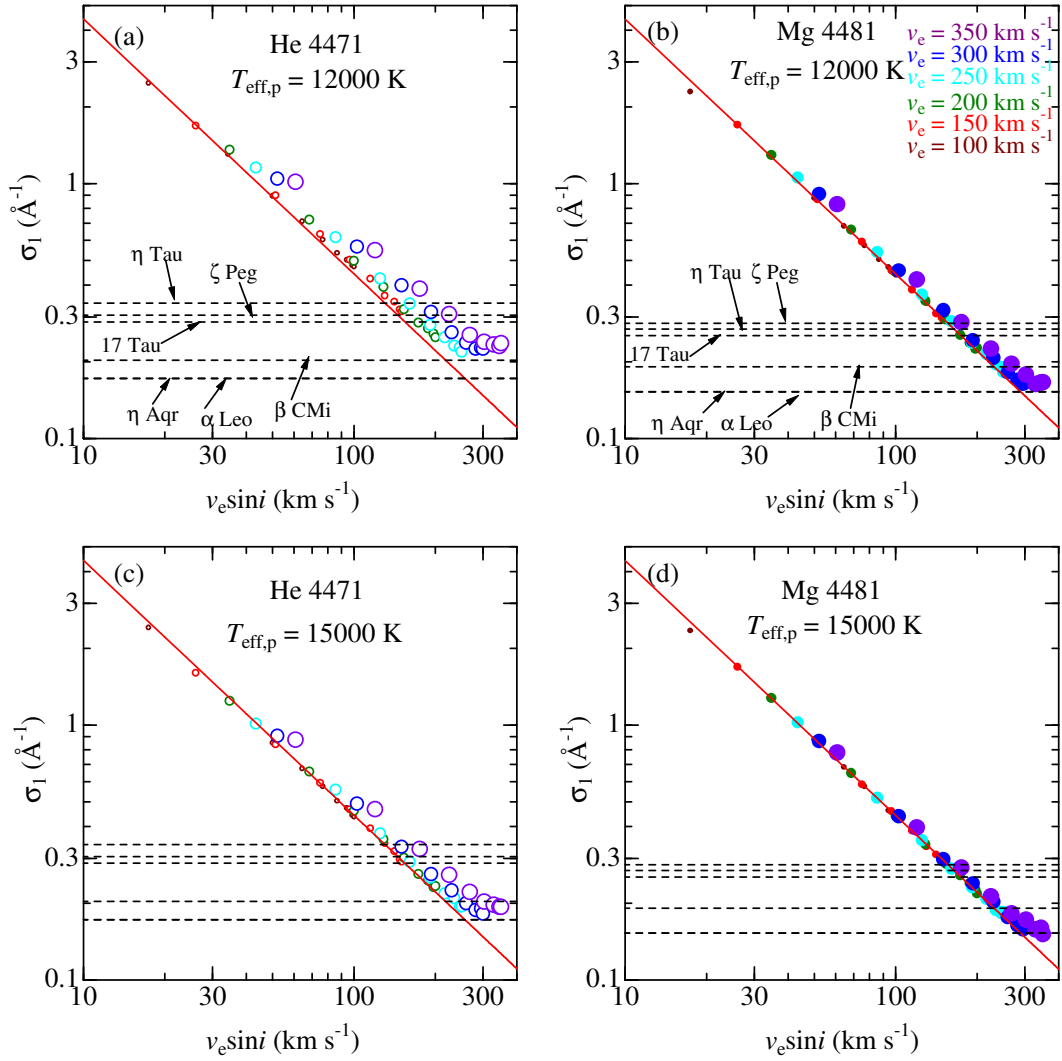}
\caption{
The values of theoretical $\sigma_{1}$, measured from the Fourier transforms of 
simulated spectra for gravity-darkened models, are plotted against $v_{\rm e}\sin i$. 
The results for He~4471 (left panels) and Mg~4481 (right panels) are shown 
in open and filled circles, respectively, where those corresponding to different 
$v_{\rm e}$ are discriminated by their size and color (larger symbol 
for higher $v_{\rm e}$). The solid line shows the classical relation
$\sigma_{1} = 0.660/(\lambda v_{\rm e}\sin i / c)$. The observed $\sigma_{1}$
values for each of the six program stars are indicated by horizontal dashed lines.
The upper panels (a) and (b) show the results for $T_{\rm eff,p}$ = 12000~K,
while the lower panels (c) and (d) are for $T_{\rm eff,p}$ = 15000~K. 
}
\label{fig7}
\end{center}
\end{minipage}
\end{figure*}

\setcounter{figure}{7}
\begin{figure*}
\begin{minipage}{110mm}
\begin{center}
\includegraphics[width=11.0cm]{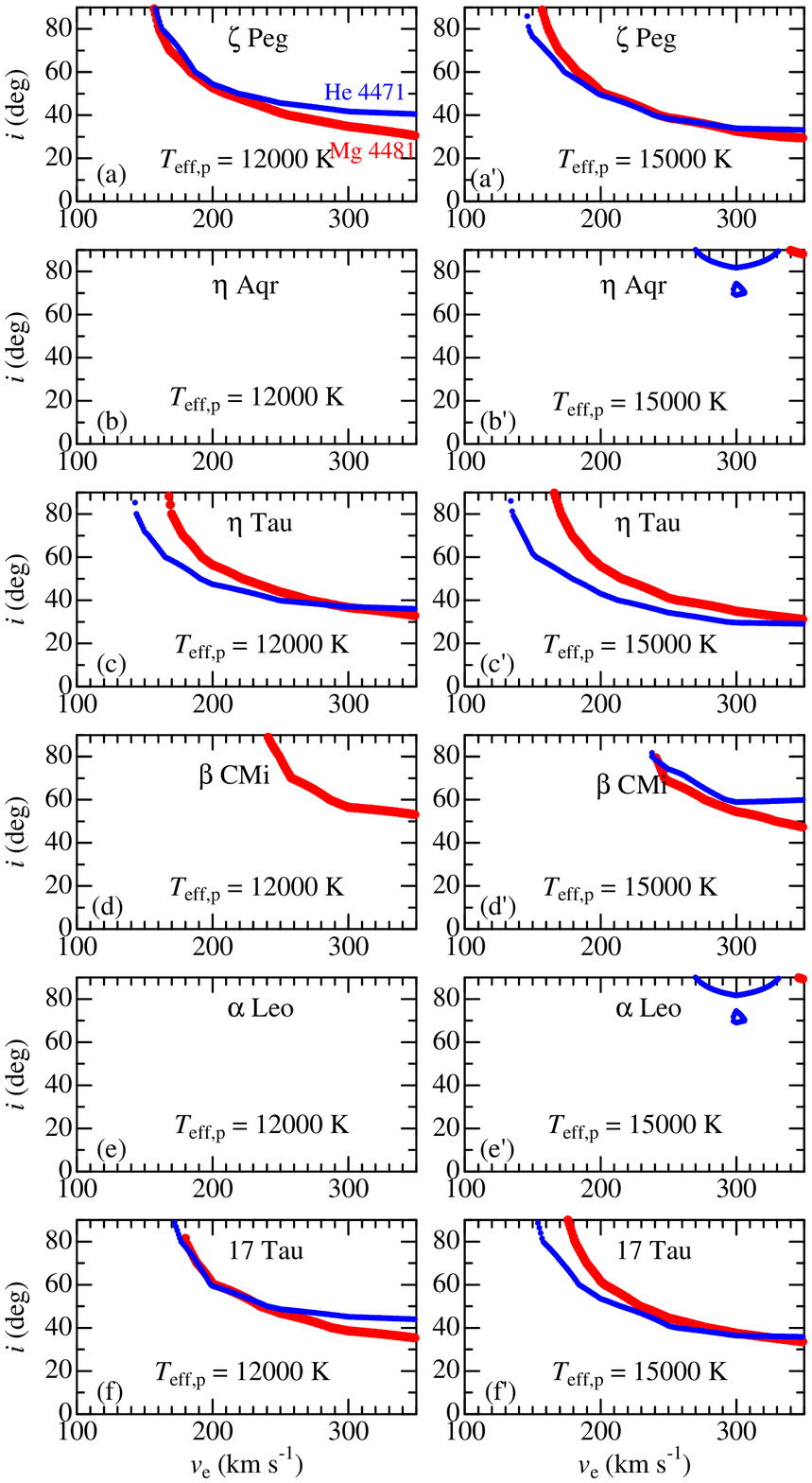}
\caption{
Graphical display of contours in the ($v_{\rm e}$, $i$) plane, which were derived as 
the solution satisfying the equation $\sigma_{1}^{\rm obs} = \sigma_{1}^{\rm th}(i,v_{\rm e})$.
The results for He~4471 and Mg~4481 are shown in thinner blue traces and thicker red traces,
respectively. From top to bottom are arranged the panels for $\zeta$~Peg, $\eta$~Aqr, 
$\eta$~Tau, $\beta$~CMi, $\alpha$~Leo, and 17~Tau, where each of the left-hand and 
right-hand panels are for $T_{\rm eff,p}$ = 12000~K and $T_{\rm eff,p}$ = 15000~K, respectively.
Note that, in deriving the results for $\eta$~Aqr and $\alpha$~Leo, the observed 
$\sigma_{1}^{\rm He}$ values for He~4471 were arbitrarily increased by 10\%, 
in order to bring the solution to a determinable level.   
}
\label{fig8}
\end{center}
\end{minipage}
\end{figure*}


\begin{thebibliography}{}
\bibitem[]{}
  Abt, H. A., Levato, H., Grosso, M., 2002, ApJ, 573, 359
\bibitem[]{}
  Arenou, F., Grenon, M., G\'{o}mez, A., 1992, A\&A, 258, 104
\bibitem[]{}
  Auer, L. H., Mihalas, D., 1973, ApJS, 25, 433
\bibitem[]{}
  Barnard, A. J., Cooper, J, Smith, E. W., 1974, JQSRT, 14, 1025
\bibitem[]{}
  Claret, A., 1998, A\&AS, 131, 395
\bibitem[]{}
  Collins, G. W., II,  Truax, R. J., 1995, ApJ, 439, 860
\bibitem[]{}
  D\'{\i}az, C. G., Gonz\'{a}lez, J. F., Levato, H., Grosso, M., 2011, 
  A\&A, 531, A143
\bibitem[]{}
  Domiciano de Souza, A., Zorec, J., Jankov, S., Vakili, F., Abe, L., 
  Janot-Pacheco, E., 2004, A\&A, 418, 781
\bibitem[]{}
  Ekstr\"{o}m, S., et al., 2012, A\&A, 537, A146
\bibitem[]{}
  ESA, 1997, The Hipparcos and Tycho Catalogues, ESA SP-1200, 
  available from NASA-ADC or CDS in a machine-readable form 
  (file name: hip\_main.dat)
\bibitem[]{}
  Espinosa Lara, F., Rieutord, M., 2011, A\&A, 533, A43
\bibitem[]{}
  Flower, P. J., 1996, ApJ, 469, 355
\bibitem[]{}
  Gigas, D., 1988, A\&A, 192, 264
\bibitem[]{}
  Gray D.~F., 2005, The Observation and Analysis of Stellar Photospheres, 3rd ed.
  (Cambridge, Cambridge University Press)
\bibitem[]{}
  Hauck, B., Mermilliod, M., 1998, A\&AS, 129, 431
\bibitem[]{}
  Hutchings, J. B., Stoeckley, T. R., 1977, PASP, 89, 19
\bibitem[]{}
  Izumiura, H., 1999, in Proc. 4th East Asian Meeting on Astronomy,
  Observational Astrophysics in Asia and its Future
  ed. P. S. Chen (Kunming: Yunnan Observatory), 77
\bibitem[]{}
  Jaschek, M., Egret, D., 1982, in Be Stars, Proc. IAU Symp. 98, 
  (eds.) M. Jaschek \& H.-G. Groth (Dordrecht: Reidel), p.261
\bibitem[]{}
  Kurucz, R. L., 1993, Kurucz CD-ROM, No. 13, ATLAS9 Stellar Atmosphere
  Program and 2 km/s Grid (Cambridge: Smithsonian Astrophysical Observatory)
\bibitem[]{}
  Levenhagen, R. S., 2014, ApJ, 797, 29
\bibitem[]{}
  McAlister, H. A., et al., 2005, ApJ, 628, 439
\bibitem[]{}
  Mihalas, D., 1972, ApJ, 177, 115
\bibitem[]{}
  Napiwotzki, R., Sch\"{o}nberner, D., Wenske, V., 1993,
  A\&A, 268, 653
\bibitem[]{}
  Przybilla, N., Butler, K., Becker, S. R., Kudritzki, R. P., 
  2001, A\&A, 369, 1009
\bibitem[]{}
  Ruusalepp, M., 1982, in Be Stars, Proc. IAU Symp. 98, 
  ed. M. Jaschek \& H.-G. Groth (Dordrecht: Reidel), p. 303
\bibitem[]{}
  Ryabchikova, T., Piskunov, N., Kurucz, R. L., Stempels, H. C., Heiter, U., 
  Pakhomov, Yu, Barklem, P.~S., 2015, Phys. Scr., 90, 054005 
\bibitem[]{}
  Sim\'{o}n-D\'{\i}az S., Herrero A., 2007, A\&A, 468, 1063
\bibitem[]{}
  Stoeckley, T. R., 1968a, MNRAS, 140, 121
\bibitem[]{}
  Stoeckley, T. R., 1968b, MNRAS, 140, 141
\bibitem[]{}
  Stoeckley, T. R., Buscombe, W., 1987, MNRAS, 227, 801
\bibitem[]{}
  Stoeckley, T. R., Mihalas, D., 1973, NCAR Technical Note, 
  NCAR-TN/STR-84
\bibitem[]{}
  Takeda, Y., 1994, PASJ, 46, 181
\bibitem[]{}
  Takeda, Y., 1995, PASJ, 47, 287
\bibitem[]{}
  Takeda, Y., Kawanomoto, S., Ohishi, N., 2008, ApJ, 678, 446
\bibitem[]{}
  Takeda, Y., Kambe, E., Sadakane, K., Masuda, S., 2010, PASJ, 62, 1239
\bibitem[]{}
  Townsend, R. H. D., Owocki, S. P., Howarth, I. D., 2004, MNRAS, 350, 189
\bibitem[]{}
  Vinicius, M.-M. F., Townsend, R. H. D., Leister, N. V., 2007, in Active OB Stars:
  Laboratories for Stellar and Circumstellar Physics, ASP Conf. Ser., Vol. 361, 
  ed. S. \u{S}tefl, S. P. Owocki \& A. T. Okazaki (San Francisco: ASP), p. 518 
\bibitem[]{}
  Zorec, J., et al., 2011, A\&A, 526, A87
\bibitem[]{}
  Zorec, J., et al., 2016, A\&A, 595, A132
\bibitem[]{}
  Zorec, J., et al., 2017, A\&A, 602, A83
\end{thebibliography}
\end{document}